\documentclass[preprint,12pt,numbers,sort&compress]{elsarticle}

\usepackage[monochrome]{color}
\usepackage{amssymb}
 \usepackage{amsthm}
\usepackage{color}
\usepackage[T1]{fontenc}
\usepackage[utf8]{inputenc} 
\usepackage{float}
\graphicspath{{fig2/}}
\usepackage{multirow}
\usepackage{subfigure}
\usepackage{graphicx}
\usepackage{mathrsfs}
\usepackage{diagbox}

\DeclareUnicodeCharacter{0301}{\'{e}}

\usepackage{easyReview}

\journal{Combustion and Flame}

\begin{document}

\begin{frontmatter}



\title{Flame interactions in a stratified swirl burner: flame stabilization, combustion instabilities and beating oscillations}


\author[buaa]{Xiao~Han} 
\author[imperial]{Davide~Laera}
\author[imperial]{Dong~Yang}
\author[buaa]{Chi~Zhang\corref{cor1}}
\author[buaa]{Jianchen~Wang}
\author[buaa]{Xin~Hui}
\author[buaa]{Yuzhen~Lin} 
\author[imperial]{Aimee S. Morgans} 
\author[uc]{Chih-Jen~Sung} 

\address[buaa]{National Key Laboratory of Science and Technology on Aero-Engine Aero-thermodynamics, Co-innovation Center for Advanced Aero-Engine, School of Energy and Power Engineering, Beihang University, Beijing, 100083, P. R. China\\}

\address[imperial]{Department of Mechanical Engineering, Imperial College London, London SW7 2AZ, UK}

\address[uc]{Department of Mechanical Engineering, University of Connecticut, Storrs, CT 06269, USA} 

\cortext[cor1]{Corresponding author: zhangchi@buaa.edu.cn}

\begin{abstract}

The present article investigates the interactions between the pilot and main flames in a novel stratified swirl burner using both experimental and numerical methods. Experiments are conducted in a test rig operating at atmospheric conditions. The system is equipped with the BASIS (Beihang Axial Swirler Independently-Stratified) burner fuelled with premixed methane-air mixtures. To illustrate the interactions between the pilot and main flames, three operating modes are studied, where the burner works with: (i) only the pilot flame, (ii) only the main flame, and (iii) the stratified flame (with both the pilot and main flames). We found that: (1) In the pilot flame mode, the flame changes from V-shape to M-shape when the main stage is switched from closed to supplying a pure air stream. Strong oscillations in the M-shape flame are found due to the dilution of the main air to the pilot methane flame. (2) In the main flame mode, the main flame is lifted off from the burner if the pilot stage is supplied with air. The temperature of the primary recirculation zone drops substantially and the unsteady heat release is intensified. (3) In the stratified flame mode, unique beating oscillations exhibiting dual closely-spaced frequencies in the pressure spectrum is found. This is observed within over narrow window of equivalence ratio combinations between the pilot and main stages. Detailed analysis of the experimental data shows that flame dynamics and thermoacoustic couplings at these two frequencies are similar to those of the unstable pilot flame and the attached main flame cases, respectively. Large Eddy Simulations (LESs) are carried out with OpenFOAM to understand the mechanisms of the time-averaged flame shapes in different operating modes. Finally, a simple acoustic analysis is proposed to understand the acoustic mode nature of the beating oscillations. 
\end{abstract}


\end{frontmatter}







\section{Introduction} 

Lean premixed prevaporized (LPP) combustors are widely used to reduce NOx emissions of modern aero-engine and land-based gas turbines. These combustors face combustion instability challenges coming from a resonant coupling between acoustic modes and unsteady heat release rate (HRR)~\cite{lieuwen2005combustion}. To broaden the stable operating range, LPP combustors are often designed as a centrally-staged structure where a pilot flame is surrounded by a main flame. These two flames are often operated with different mass flow rates ($\dot{m}_a$) and equivalence ratios ($\phi$)~\cite{dhanuka2009vortex},  leading to a local equivalence ratio difference when they interact with each other. This mixture inhomogeneity forms a so-called stratified flame that has attracted many studies~\cite{nogenmyr2007large,kuenne2012experimental}. Usually, a richer and stable pilot flame is preferable to stabilise a leaner main flame and thus reduce emissions~\cite{mongia2003challenges}. These two flames are usually separated by a lip structure, as sketched in Fig.~\ref{rig1}, to delicately adjust the mixing and merging of the two streams -- this has a direct impact on the entire flow field and equivalence ratio distribution and has been shown to play a key role in NOx reduction~\cite{li2016emission}.

There are some recent studies on structures and dynamics of stratified swirl flames in centrally-staged combustors. For example, in a non-separated stratified swirl burner, Chong \emph{et al.}~\cite{chong2016effect} observed that with a larger mass flow rate in the pilot stream, the flame has a longer and stronger reaction zone. Kim and Hochgreb~\cite{kim2011nonlinear} found that the split of equivalence ratios between the two stages plays a key role in determining the flame shape and the flame dynamics. The flame response of the above stratified flames and the spatial distribution of heat release were then measured experimentally by Han \emph{et al.}~\cite{han2015response,han2015spatial}. In a staged dual swirl burner with liquid fuel, Renaud \emph{et al.}~\cite{renaud2017bistable} observed bistable behaviour and flame shape transition, which can be triggered by acoustic perturbation. The above two burners do not feature a separation structure, therefore the pilot and main flames often blend with each other. Using industrial LPP combustors, self-excited limit cycle oscillation operated at elevated-pressure conditions was measured and analyzed by Temme \emph{et al.}~\cite{temme2014combustion}. Li \emph{et al.}~\cite{li2016emission} found that the equivalence ratios affect the combustion efficiency and emissions significantly in a high-pressure configuration. But both combustors in Refs~\cite{temme2014combustion,li2016emission} feature complicated geometries and were not shown in detail. Recently, a simplified stratified burner was developed by the authors' group~\cite{han2019flame} to elucidate the role of the pilot flame in anchoring the main flame experimentally.

From these studies, it has been shown that interactions between the pilot and the main flames play an important role in the dynamics of stratified swirl flames. However, to the best knowledge of the authors, none of the previous works on combustion instabilities in centrally-staged combustors have studied the characteristics of each of these two swirl flames and their interactions in detail. Chen \emph{et al.}~\cite{chen2019interaction} performed Large Eddy Simulations (LESs) in a dual-swirl burner without fuel-staging, finding that, due to their acoustic impedance difference, the velocity perturbations at the two air passages are different during self-excited oscillations, leading to the flapping of the fuel jets and radial equivalence ratio fluctuations. Recently, Kim et at.~\cite{kim2019experimental} reported interesting beating oscillations in a stratified swirl burner without lip structures, with a slight difference in oscillating frequencies between the pilot and the main flames. This beating phenomenon may shed light on how flame interactions between the pilot and the main stages could affect combustion instabilities in centrally-staged combustors. However, no detailed measurements and analysis were provided and thus physical insights of the flame dynamics and flame-acoustic interaction mechanisms are still missing. 

Even in simple configurations, such as a Rijke tube or a single swirler burner, many complex nonlinear behaviours have been widely observed, e.g. see Refs.~\cite{Kashinath2014a,kabiraj2012nonlinear,durox2009experimental,balusamy2015nonlinear}. A recent comprehensive review can be found in~\cite{juniper2018sensitivity}. Among these cases, beating comes in as an oscillation phenomenon characterised by a moduled periodic pressure fluctuation with a low-frequency envelope in its amplitude. Mathematically, it comes from the linear superposition of two periodic signals with close frequencies and similar amplitudes:
\begin{equation}
 P= A_1\cos(2\pi f_1+\psi_1) + A_2\cos(2\pi f_2+\psi_2),
\label{eq_beating}
\end{equation}
where $A$, $f$, and $\psi$ are the amplitude, frequency, and phase shift of each signal, with the frequency of the wave package being the difference between $f_1$ and $f_2$. Choi \emph{et al.}~\cite{choi2005control} measured the flame structure during the beating oscillations of a premixed swirl flame and found that the flame is highly distorted near the medium perturbation amplitude. Beating oscillations were also found by Kabiraj and Sujith~\cite{kabiraj2012nonlinear} with the elongation and pinch-off of a laminar premixed conical flame. Weng \emph{et al.}~\cite{weng2016investigation} investigated beating oscillations in a Rijke burner, showing that flame heat loss can play an important role in the low-frequency flame pulsation. Moon \emph{et al.}~\cite{moon2019combustion} observed beating oscillations in adjacent model combustors with cross-talk area. However, these beating phenomena come from either nonlinear behaviours of a single flame or cross-talk between different burners. The effect of flame interactions between the pilot and main stages on affecting the beating phenomena in centrally-staged combustors is relevant to many modern gas turbine applications but remains unclear. 

In the present study, we aim to investigate the effect of pilot-main flame interactions on flame stabilization, flame shape, flame dynamics and acoustic-flame interactions during self-excited combustion instabilities within a stratified swirl burner. Three operating modes are studied: pilot flame mode (fuel comes from only the pilot stage), main flame mode (fuel comes from only the main stage), and stratified flame mode (fuel comes from both the pilot and main stages). Both experiments and LES are performed. The paper is organized as follows: the experimental and numerical setups are introduced first, each flame mode is then studied in detail -- including their time-averaged flame shapes, LES results, pressure fluctuations, flame dynamics and acoustic-flame couplings, and a simplified acoustic analysis is proposed at last to qualitatively illustrate the mechanism of the observed beating oscillations.

\section{Experimental and numerical setup}
\subsection{Experimental setup}
\begin{figure}[htbp]
\centering
     \subfigure[\label{rig1}]{%
       \includegraphics[width=0.5\textwidth]{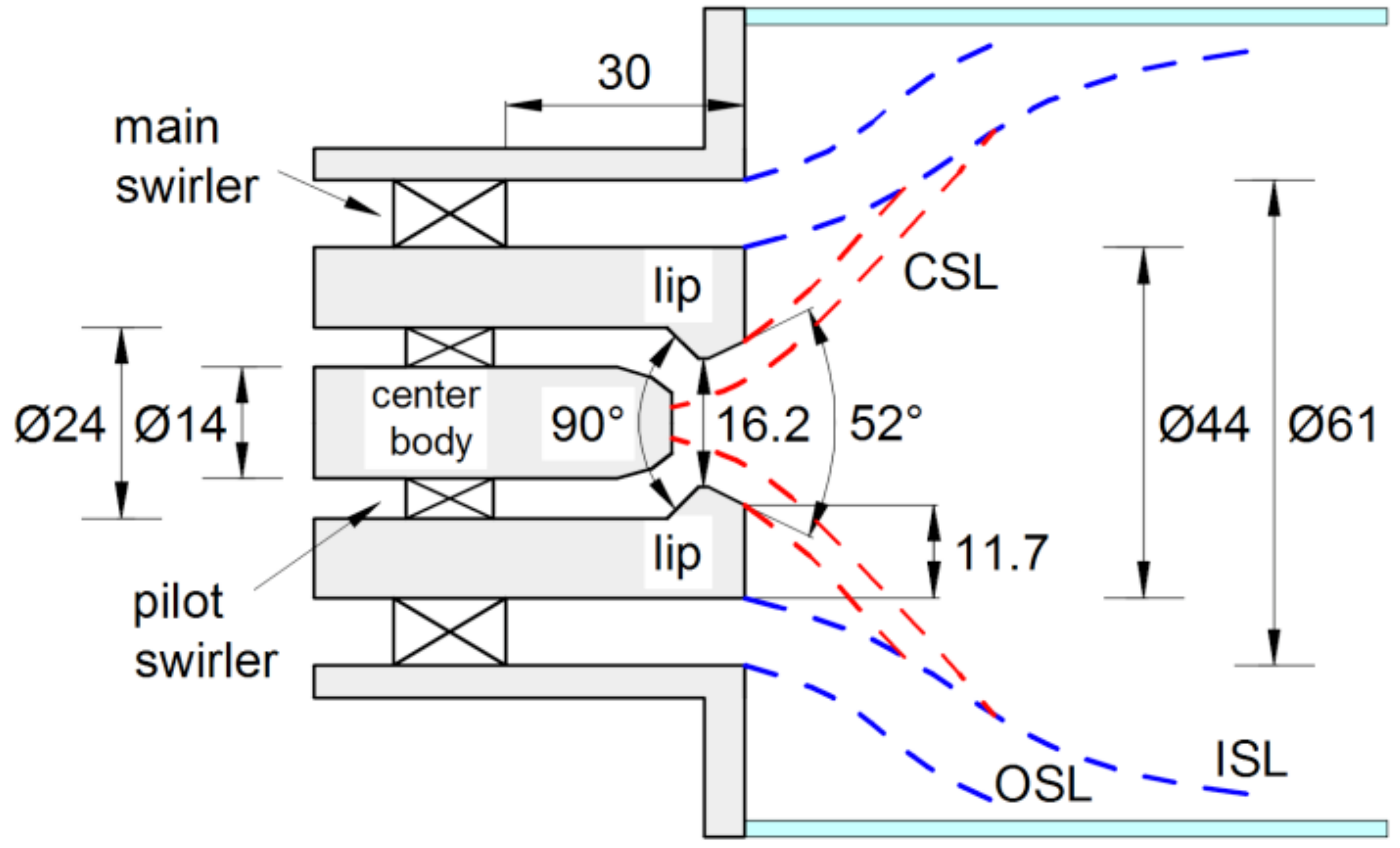}
     }\\
     \subfigure[\label{rig2}]{%
       \includegraphics[width=0.75\textwidth]{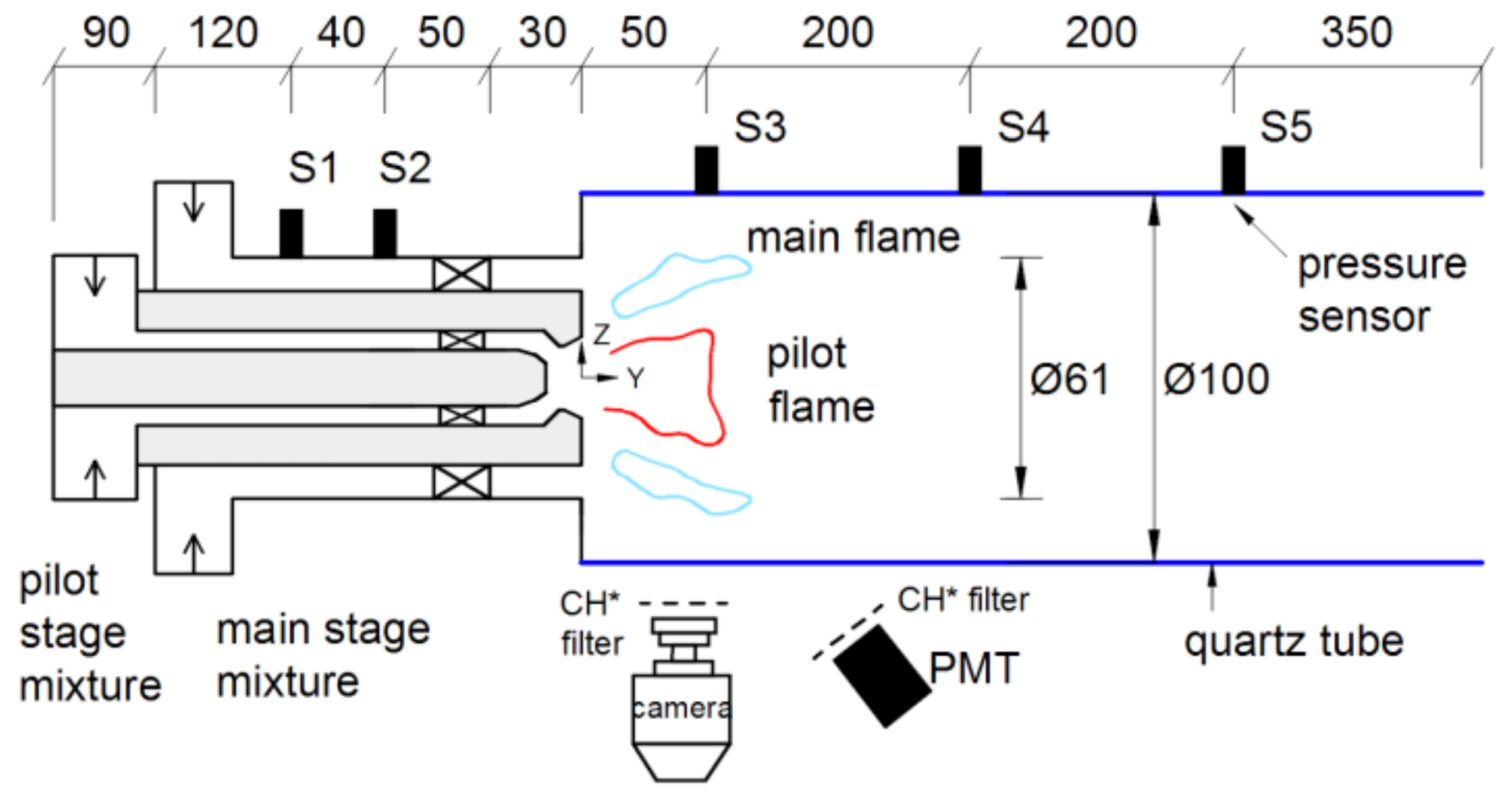}
     }\\
\caption{(a) A schematic view of the BASIS burner. CSL stands for the central shear layer originating from the pilot stage, while ISL and OSL stand for the inner and outer shear layer of the main stage, respectively. (b) A schematic view of the test rig (not to scale)~\cite{han2019flame}. All dimensions are in millimetre.} 
\label{rig} 
\end{figure}

In the present work, a novel burner (BASIS) designed at Beihang University~\cite{han2019flame,han2018effect} is used to study the aero-thermodynamic behaviour and combustion instabilities of LPP combustors. Its detailed geometry is sketched in Fig.~\ref{rig1}. Two co-axial co-rotation axial swirlers are used -- one for the pilot stage and the other for the main stage, with swirl numbers 0.68 and 0.50, respectively. 

Figure~\ref{rig2} shows the setup of the test rig. The burner is operated at atmospheric pressure with fully premixed methane-air mixtures. The burner is connected to a quartz tube of 100 mm in diameter. Two quartz tubes with different lengths are chosen. The shorter one has a length of 200 mm and is employed only to measure the time-averaged flame shape, without triggering combustion instabilities. The longer tube has a length of 800 mm and is used to investigate combustion instabilities.

During the experiments, the mass flow rates of air are monitored by standard orifice flowmeters (accuracy of 2 \%), while those for the methane are regulated by two Mass Flow Controllers (Sevenstar, CS200) (accuracy of 1 \%). A high-speed camera (Photron, Fastcam SA4) equipped with an intensifier is used to record the flame dynamics. The view window is set as 648 $\times$ 768 pixels, corresponding to an actual dimension of 84 mm $\times$ 100 mm. 3882 snapshots are obtained for each case with a sampling rate of 5000 fps. A digital single-lens reflex (DSLR) camera is used to capture the time-averaged flame shape with an aperture of f/10, exposure time of 1/2 s, and ISO value of 640. The global CH$ ^* $ chemiluminescence ($I'$) is measured by a photomultiplier (PMT) (Hamamatsu, H9306) as a representative of HRR~\cite{kim2011nonlinear}. CH$ ^* $ filters (430$ \pm $5 nm) are equipped to the above three optical instruments. Five pressure sensors (PCB, 112A22) are mounted along with the test rig to measure the pressure fluctuation $P'$, as shown in Fig.~\ref{rig2}. Note that S1 and S2 are flush mounted to the main channel, while S3-S5 are recess mounted with semi-infinite tubes due to high temperature. All the above data are synchronised and collected by a DAQ system (National Instruments, NI9215). In each case, signals of $I'$ and $P'$ are recorded at a sampling frequency of 20 kHz for 5 seconds.

\begin{table}[htbp]
\caption{Operating conditions of the current testing.}
\begin{center}
\label{cond}
\begin{tabular}{cccccccccccc}
\hline
Case &  Operating mode  &\begin{tabular}[c]{@{}c@{}}$\dot{m_a}_p$\\ (g/s)\end{tabular}  & $\phi_p$ & \begin{tabular}[c]{@{}c@{}}$\dot{m_a}_m$\\ (g/s)\end{tabular}  & $\phi_m$ & $\phi_{total}$ & \begin{tabular}[c]{@{}c@{}}$P$\\ (kW)\end{tabular}       \\\hline
$P_fM_c$   &  Pilot flame + Main closed  & 2.2             & 0.85       & 0               & 0          & 0.85       & 6.1          \\
$P_fM_a$  & Pilot flame + Main air & 2.2             & 0.85       & 17.8            & 0          & 0.09       & 6.1          \\
$P_cM_f$  & Pilot closed + Main flame & 0               & 0          & 17.8            & 0.63       & 0.63       & 36.4    \\   
$P_aM_f$  & Pilot air + Main flame & 2.2             & 0          & 17.8            & 0.63       & 0.56       & 36.4      \\
$P_fM_f$   & Pilot flame + Main flame & 2.2             & 0.85       & 17.8            & 0.63       & 0.65       & 42.5            \\
\hline
\end{tabular}
\end{center}
\end{table}

The test conditions are chosen as follows: the air mass flow rates for the pilot and main stages are kept at 2.2 g/s and 17.8 g/s, respectively. This results in an air split ratio of about 1:8, being consistent with the typical split in LPP combustors. The mean axial velocities of in the pilot and main channels are about 6.5 m/s and 11 m/s, resulting in Reynolds numbers of approximate 4000 and 12000, respectively. 

Five cases are chosen to study interactions between the pilot and main stages, as listed in Table~\ref{cond}. The capital letters $P$ and $M$ stand for the pilot and main stages, respectively, while the subscripts $f$, $a$ and $c$ denote supplied with fuel, supplied with air and stage closed, respectively. Cases $P_fM_c$ and $P_fM_a$ are operated with the pilot flame where the fuel is provided only in the pilot stage (in the form of the premixed mixture). The difference is that $P_fM_c$ works with only the pilot stream (the main stage is closed), while in $P_fM_a$ the main stage is open but supplies only air rather than a fuel-air mixture. Cases $P_cM_f$ and $P_aM_f$ follow the same rules except that the fuel is now provided only in the main stage. Case $P_fM_f$ is the stratified flame case, i.e., fuels are provided in both the pilot and main stages.

\subsection{Numerical methods}
To investigate the time-averaged flow fields and reveal the flame stabilization mechanisms of different cases, LES using OpenFOAM\footnote{https://openfoam.org}~\cite{weller1998tensorial} is performed with the incompressible solver ReactingFoam. With the approximation that density is only a function of the temperature, an incompressible solver can capture the time-averaged flame shapes and flow fields, as validated previously~\cite{han2018effect,han2015openfoam, xia2019numerical}. The WALE sub-grid model~\cite{nicoud1999subgrid} is applied to close the Favre-filtered reactive flow conservation equations. The partially-stirred reactor (PaSR) combustion model~\cite{sabelnikov2013combustion} is used to model the filtered chemical reaction rates based on a global four-step methane-air reaction mechanism with six intermediate species~\cite{jones1988global}. The computational domain begins 55 mm upstream of the dump plane (11 mm upstream of the outer swirler) and ends 200 mm downstream, and is discretised using a 10-million-elements structured mesh. The near-wall cells in the flame tube are locally refined to meet the y$^+$<2 requirement. Please refer to Fig. A in Supplemental Material. A non-slip boundary condition is applied at all the solid walls. The thermal boundary condition is set by applying a different wall temperature to account for the heat loss effect~\cite{fiorina2015challenging,bauerheim2015sensitivity} with a constant 700 K imposed at the dump plane, and a linear distribution from 700 K to 1500 K fixed on the side wall of the flame tube, these values being derived from experimental observations. All the other walls are adiabatic. This approach has been validated in Refs.~\cite{han2018effect,benard2019large}. The Courant number is limited to below 0.5. The presented time-averaged flow field is based on about 5 flow through cycles after transients have died away. The validation of LES method can be found in Supplemental Material (\emph{cf.} Figs. B and C). The employed numerical setup has been proven being able to predict experimentally observed time-averaged flame shapes and hydrodynamic flame dynamics in the present stratified swirl burner under different test conditions~\cite{han2018effect}. 

\section{Results and discussion}
The experimental results are briefly summarised in Table~\ref{results}, including the flame shapes and thermoacoustic stabilities. It is found that only Case $P_fM_c$ exhibits stable combustion while the other four cases show large amplitude oscillations. A dual-frequency distribution is found in the pressure spectrum of Case $P_fM_f$ as beating oscillations~\cite{kim2019experimental,weng2016investigation}. Detailed results and discussion of these three flame modes will be presented hereafter. 

\begin{table}[htbp]
\caption{Tested cases with indication of the flame shape.}
\begin{center}
\label{results}
\begin{tabular}{ccccccccccr}
\hline
Case &           Flame shape    & Stability                   & $f$ (Hz) & $P^{\prime}$ (Pa) \\\hline
$P_fM_c$          &  V-shape    & stable & -  & -             \\
$P_fM_a$       &  M-shape & unstable                  & 159          & 1057           \\
$P_cM_f$  &  M-shape       & unstable                  & 222          & 143        \\   
$P_aM_f$     &  Lifted       & unstable                  & 224            & 466            \\
$P_fM_f$    &  S-shape       & beating                      & 198/223          & 144/128            \\
\hline
\end{tabular}
\end{center}
\end{table}

\subsection{Pilot flame mode}
 \label{Pilot flame mode}
\begin{figure}[htbp]
\centering
\includegraphics[width=0.4\textwidth]{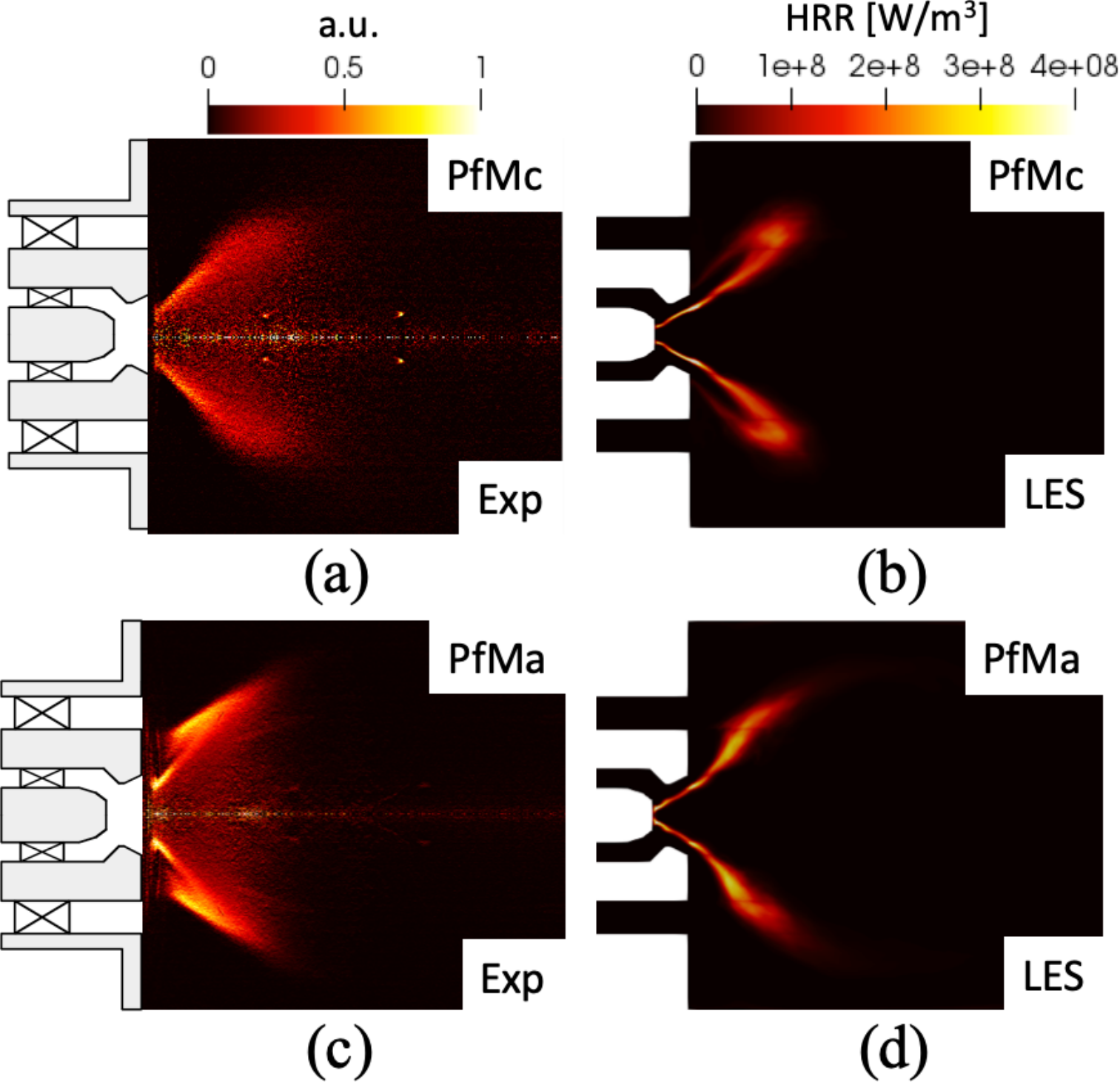}
\caption{\textcolor{red}{Time-averaged flame shapes captured by the DSLR camera (a, c) and LES HRR (b, d) for Cases $P_fM_c$ and $P_fM_a$.}}
\label{les_pilot_1} 
\end{figure}

The time-averaged flame shapes of Cases $P_fM_c$ and $P_fM_a$ are shown in \textcolor{red}{Fig.~\ref{les_pilot_1}. For both cases, measurements (left) are found in good agreement with LES HRR (right).} It is found that Case $P_fM_c$ \textcolor{red}{(Fig.~\ref{les_pilot_1}(a-b))} features a V-shape flame with thin flame sheets, and the flame anchors only at the centre body of the pilot stage. On the contrary, Case $P_fM_a$ \textcolor{red}{(Fig.~\ref{les_pilot_1}(c-d))}  sees a more-expanded M-shape flame with two stagnation points, one at the centre body and the other at the lip structure. The flame sheets of Case $P_fM_a$ are thicker, indicating that the shear layer is more turbulent than $P_fM_c$.

\begin{figure}[htbp]
\centering
\includegraphics[width=0.6\textwidth]{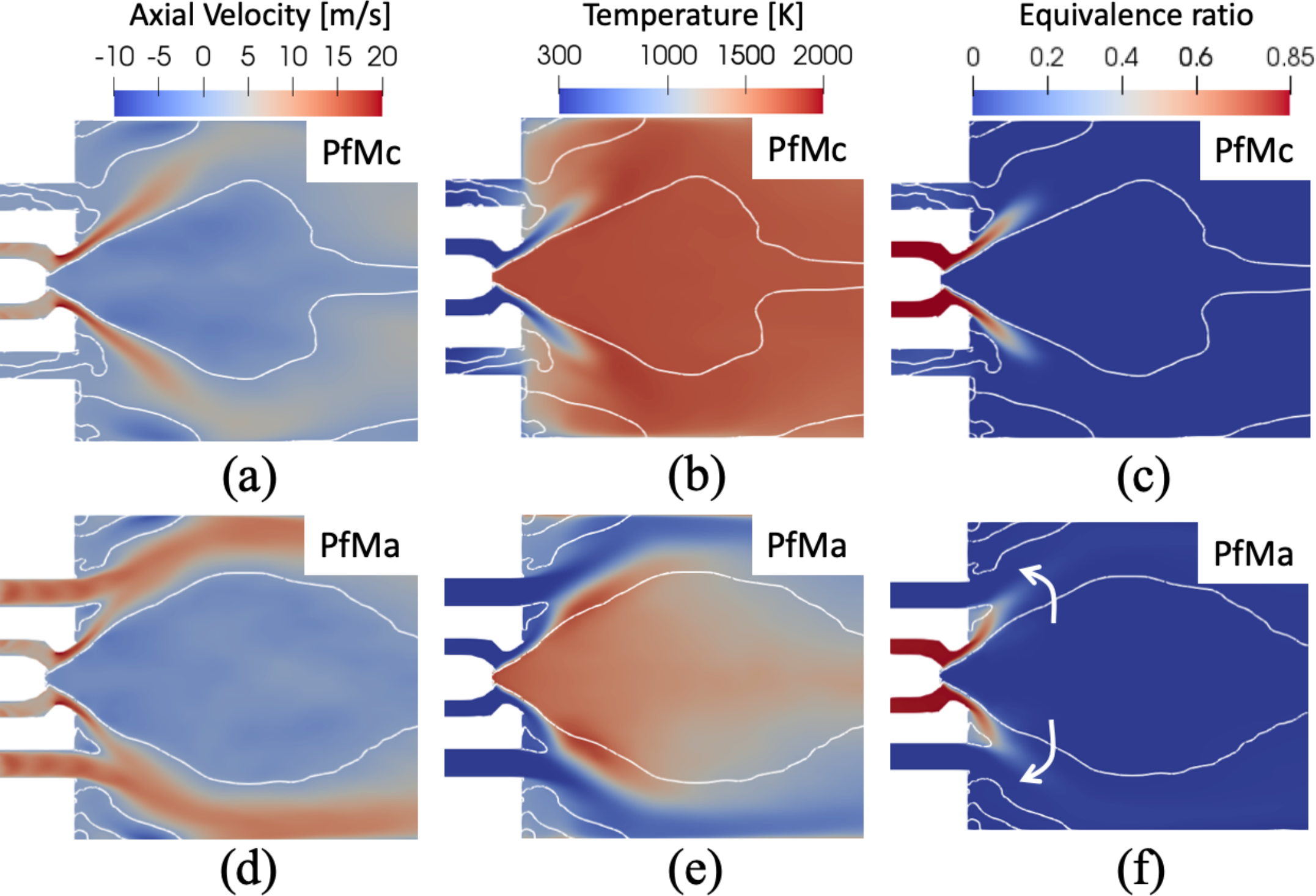}
\caption{\textcolor{red}{Time-averaged LES results for Cases $P_fM_c$ and $P_fM_a$. The distributions of axial velocity (a, d), temperature (b, e), and equivalence ratio (c, f) are shown respectively. White lines mark the zero contour of axial velocity.}}
\label{les_pilot_2} 
\end{figure}

An explanation for the above differences is provided by analysing the LES results shown in Fig.~\ref{les_pilot_2}. In Case $P_fM_c$, the pilot flow is not subject to interference as the main stage is closed. Indeed, it forms a nearly straight jet (Fig.~\ref{les_pilot_2}(a)) with the jet/flame angle dominated by the diverging angle of the pilot stage (52$^\circ$ in Fig.~\ref{rig1}) and swirl effect. The equivalence ratio distribution in Fig.~\ref{les_pilot_2}(c) exhibits a concentration along the jet, confirming the above findings. Contrarily, in Case $P_fM_a$ the main stage is supplied with an air stream at 11 m/s. This introduces a stronger shear effect, and forces the pilot stream to expand more and eventually merge with the main jet, as shown in Fig.~\ref{les_pilot_2}(d). Therefore, methane from the pilot stage enters the lip recirculation zone (LRZ) and is then diluted by the main stream, as marked by the white arrows in Fig.~\ref{les_pilot_2}(f), leading to the formation of the observed thicker M-shape flame. \textcolor{red}{The temperature maps in the two cases are also quite different. In Case $P_fM_c$ (Fig.~\ref{les_pilot_2}(b)), hot gasses occupy the entire flame tube, leading to a higher mean combustion chamber temperature than Case $P_fM_a$ (Fig.~\ref{les_pilot_2}(e)).}

\begin{figure}[htbp]
\centering
     \subfigure[\label{limit-p1}]{%
       \includegraphics[width=0.48\textwidth]{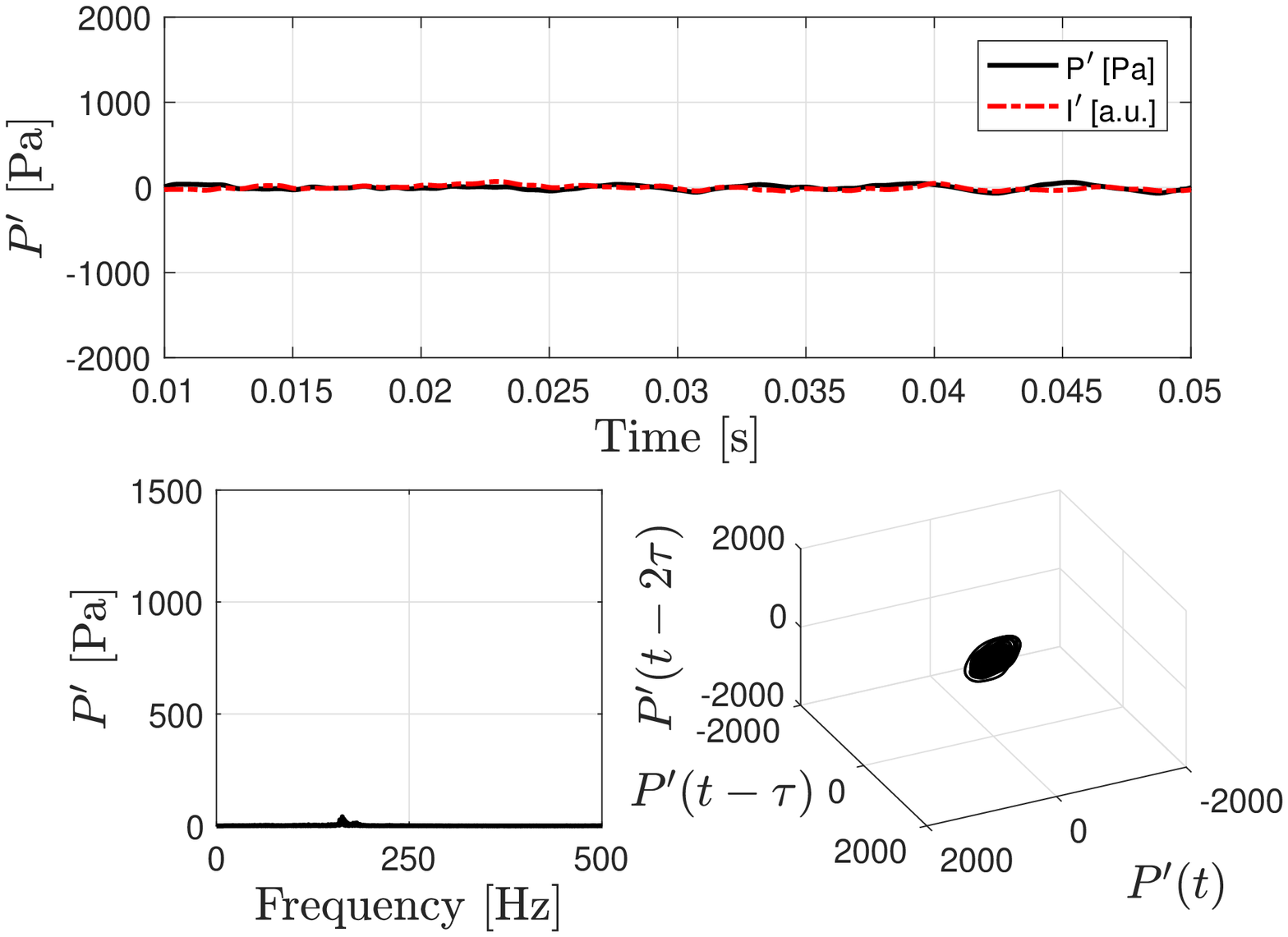}
     }
     \subfigure[\label{limit-p2}]{%
       \includegraphics[width=0.48\textwidth]{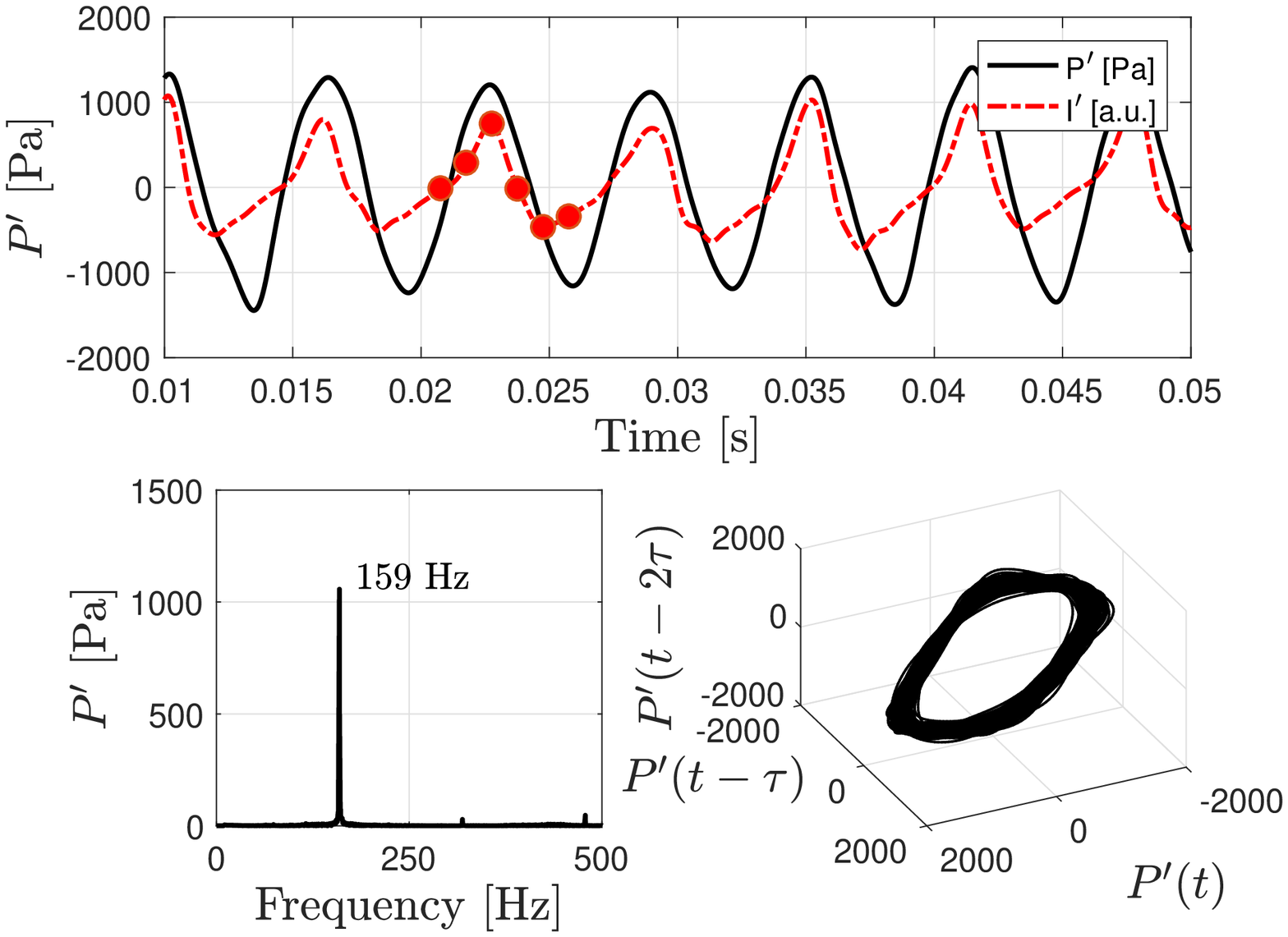}
     }\\
          \subfigure[\label{flame-p2}]{%
       \includegraphics[height=0.2\textwidth]{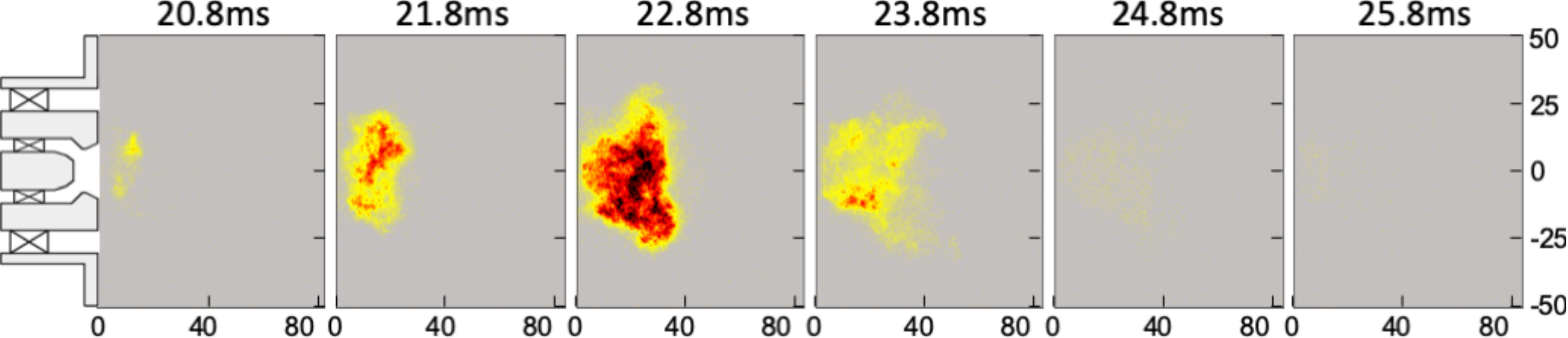}
       }\\
\caption{Pressure fluctuations and flame dynamics of the pilot flame mode. Comparison of pressure fluctuations of S3 sensor for Cases $P_fM_c$ (a) and $P_fM_a$ (b). The time-series is shown on the top, while the spectra and the phase space trajectories are shown in the bottom left and right, respectively. (c) Sequential flame images of Case $P_fM_a$ with corresponding locations in the time signal shown in (b) by red dots.}
\label{limit_pilot} 
\end{figure}

The pressures recorded by sensor S3 are used to study the system thermoacoustic behaviours. The comparison of Cases $P_fM_c$ and $P_fM_a$ is shown in Fig.~\ref{limit_pilot}. Time-series of pressure fluctuation $P'$ and flame intensity fluctuation $I'$ (from PMT) are plotted together on the top. The spectra are obtained with the Fast Fourier Transform (FFT) method, while the phase space reconstructions follow the method stated in~\cite{han2019inlet}. Case $P_fM_c$ is thermoacoustically stable, as the pressure fluctuation signal shown in Fig.~\ref{limit-p1} sees a flat spectrum and a chaotic phase trajectory. Contrarily, a strong periodic oscillation is found in Case $P_fM_a$(Fig.~\ref{limit-p2}) where $I'$ is almost synchronous with $P'$, resulting in a high single peak in the spectrum at 159 Hz with amplitude over 1000 Pa. The analysis of the frequency will be shown in Section~\ref{sub:acoustics}.

The sequential flame images within one cycle are shown in Fig.~\ref{flame-p2}. Starting from a tiny kernel at 20.8 ms, the flame grows to a trapezoidal shape (21.8 ms). Then it reaches its largest size (with the highest flame intensity) at 22.8 ms where a curly flame tip is noticeable -- this is most probably caused by the vortex roll-up~\cite{palies2010combined}. After that, the flame shrinks back to the exit of the pilot stage, becomes almost invisible at 24.8-25.8 ms and appears again in the next cycle as what happens at 20.8 ms. Overall, the flame features a periodic expansion and contraction in fluctuations of both flame area and heat release intensity, but without significant convective effect in space. To differentiate this from convective motions in later cases, we call it ``bulk oscillation''. This bulk oscillation behaviour is often experienced in partially premixed~\cite{stohr2017interaction,wang2019combustion} and spray ~\cite{temme2014combustion,han2019inlet} flames. It is driven by both velocity and equivalence ratio fluctuations, resulting in both flame area and heat release intensity modulations. However, for a premixed flame, as its equivalence ratio is fixed, flame area modulations, such as surface wrinkling or vortex roll-up, dominate the flame dynamics when it is exposed to velocity fluctuations~\cite{durox2009experimental,stohr2017interaction,steinbacher2019consequences}. In the present configuration, as shown in Fig.~\ref{les_pilot_2}, the mixture from the pilot stream is diluted by the main stream, leading to very lean combustion (closer to the lean blowout limit) which is susceptible to instability~\cite{lieuwen1998role}. Furthermore, with the main stage air, the premixed pilot stream is turned into a partially premixed combustion regime, where the unsteady mixing between the two streams intensifies the perturbation of the equivalence ratio.

\begin{figure}[htbp]
\centering
     \subfigure[\label{dmd-p2}]{%
       \includegraphics[width=1\textwidth]{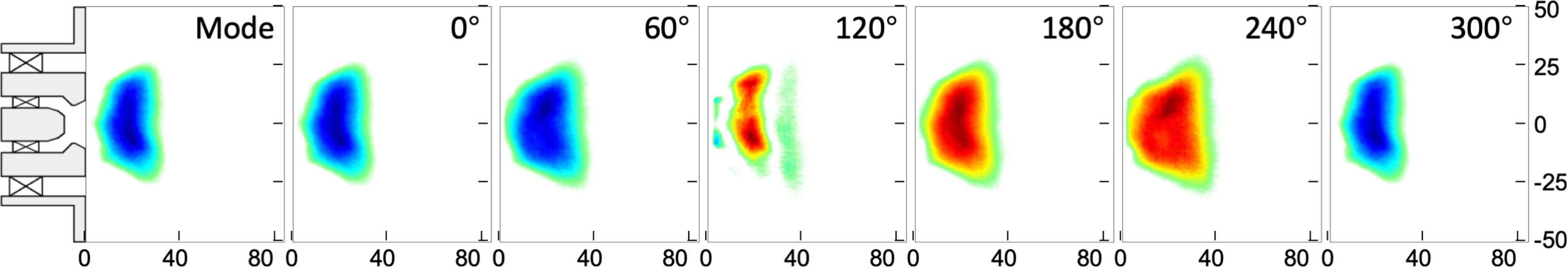}
     }\\
     \subfigure[\label{rayleigh_p2}]{%
       \includegraphics[width=0.3\textwidth]{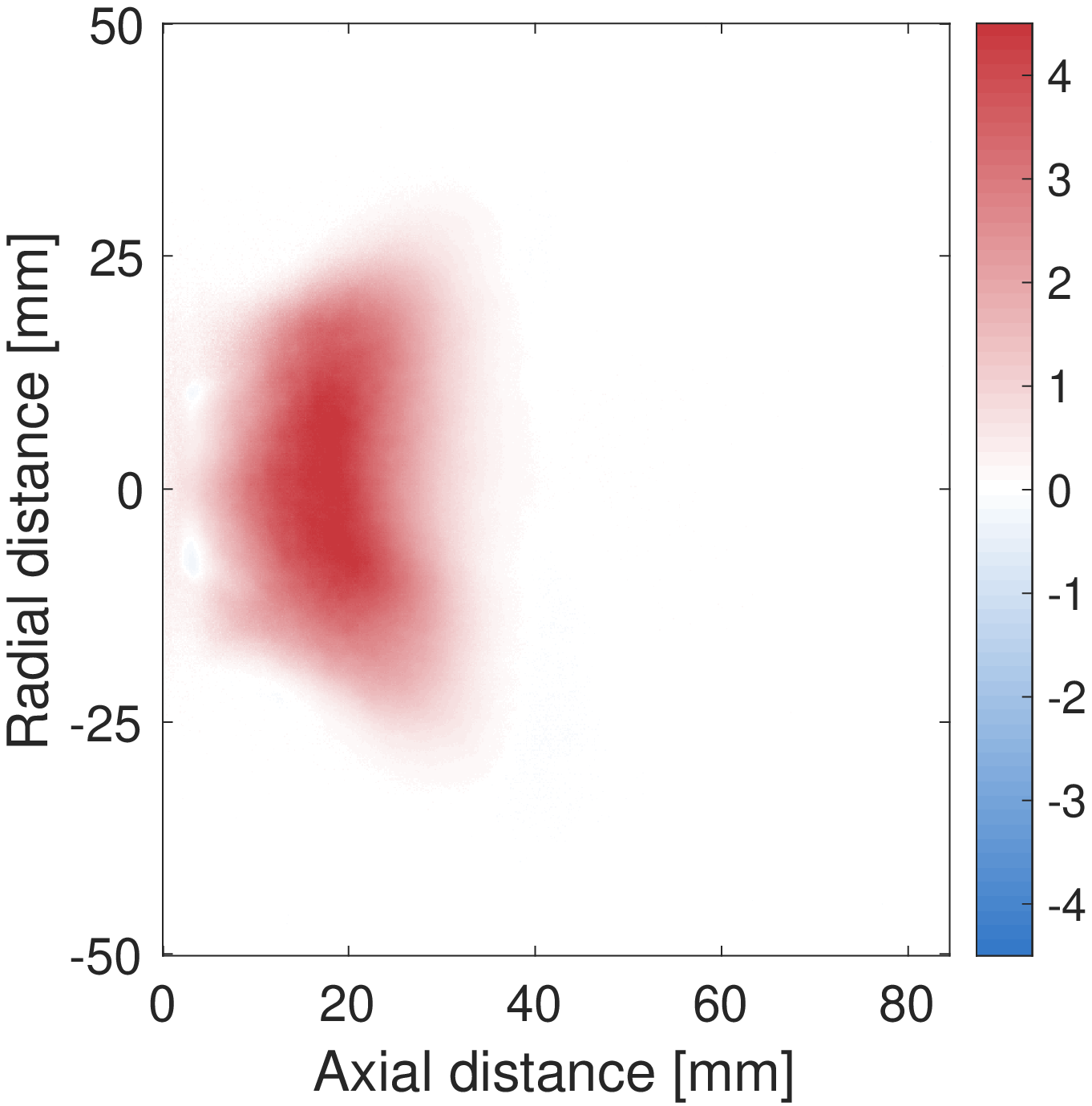}
     }\\
\caption{DMD modes (a) and local Rayleigh index map (b) for Case $P_fM_a$.}
\label{dynamics_p2}
\end{figure}

More insights of flame dynamics are given via a Dynamic Mode Decomposition (DMD)~\cite{schmid2010dynamic} of the flame images, which extracts the coherent structures during the oscillations. In Fig.~\ref{dmd-p2}, the DMD mode is plotted at the very left, with the phase sequences on its right. This result also confirms the bulk motion of Case $P_fM_a$ flame, as the axial location of the mode is almost fixed and features only area changes.

A more quantitative analysis of the thermoacoustic coupling can be illustrated by the local Rayleigh index map. This index map can be calculated by $R_{local}=\tau^{-1}\int_{\tau}^{ }P'\cdot I_{CH^*}'dt$ for each pixel of the measured domain~\cite{temme2014combustion}, where $\tau$ is the chosen processing period, $I_{CH^*}'$ represents the HRR fluctuation, using the CH$^*$ chemiluminescence recorded by the high-speed camera. For each case, a period of 60 cycles is chosen so $\tau\approx 0.38$ s. According to the Rayleigh criterion~\cite{rayleigh1878explanation}, a positive (negative) value stands for a constructive (destructive) effect driving (damping) the thermoacoustic instabilities. The Rayleigh index map of Case $P_fM_a$ is shown in Fig.~\ref{rayleigh_p2} where the whole flame is positive, indicating that the flame heat release perturbation in the whole region is positively coupled with $P'$. The dynamics of $P_fM_a$ flame does not see noticeable phase shift in space. This is in consistency with the bulk motion observed in Fig.~\ref{flame-p2} and \ref{dmd-p2}, that the pilot flame features global expansion-contraction without obvious convection in space, which is similar to that of a partially premixed flame~\cite{stohr2017interaction,wang2019combustion}.

\subsection{Main flame mode}

\begin{figure}[htbp]
\centering
\includegraphics[width=0.4\textwidth]{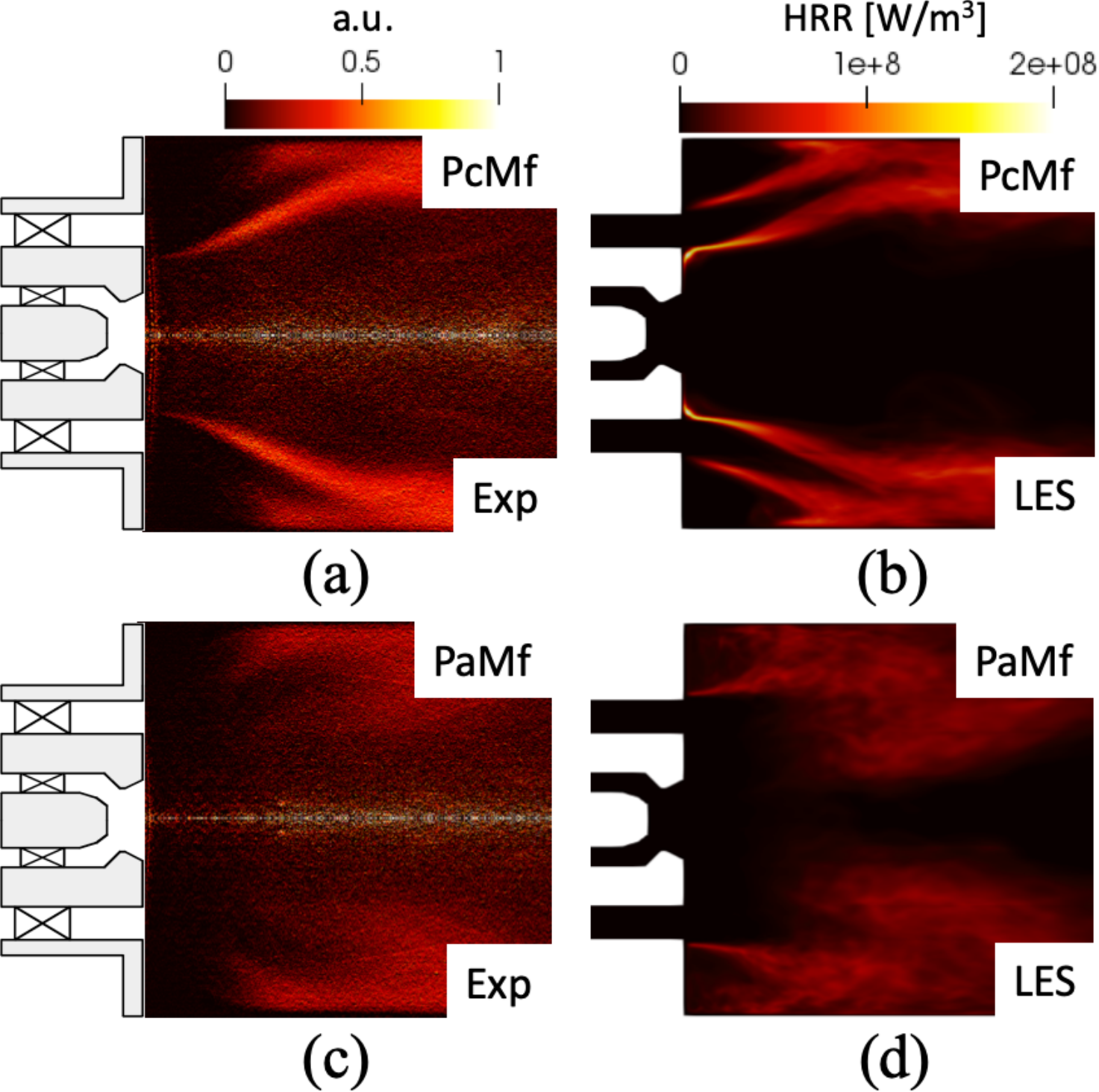}
\caption{\textcolor{red}{Time-averaged flame shapes captured by the DSLR camera (a, c) and LES HRR (b, d) for Cases $P_cM_f$ and $P_aM_f$, respectively.}}
\label{les_main_1} 
\end{figure}

\textcolor{red}{The comparison between the experimental and numerical time-averaged flame shapes for Cases $P_cM_f$ and $P_aM_f$ is shown in Fig.~\ref{les_main_1}. In Case $P_cM_f$ (Fig.~\ref{les_main_1}(a)), a V-shape flame attached only to the edge of the main stage is captured by the DSLR camera}. This flame expands at a 50$^\circ$ angle approximately, then impinges on the side wall. One can notice that the LES captures the main experimental flame shape in Fig.~\ref{les_main_1}(b), with a mismatch at the OSL most probably due to an underestimated heat loss from the wall. \textcolor{red}{The flame in Case $P_aM_f$ is instead} lifted far from the burner, as shown in Fig.~\ref{les_main_1}(c). This lift-off phenomena is well reproduced by LES \textcolor{red}{(Fig.~\ref{les_main_1}(d))}.

\begin{figure}[htbp]
\centering
\includegraphics[width=0.6\textwidth]{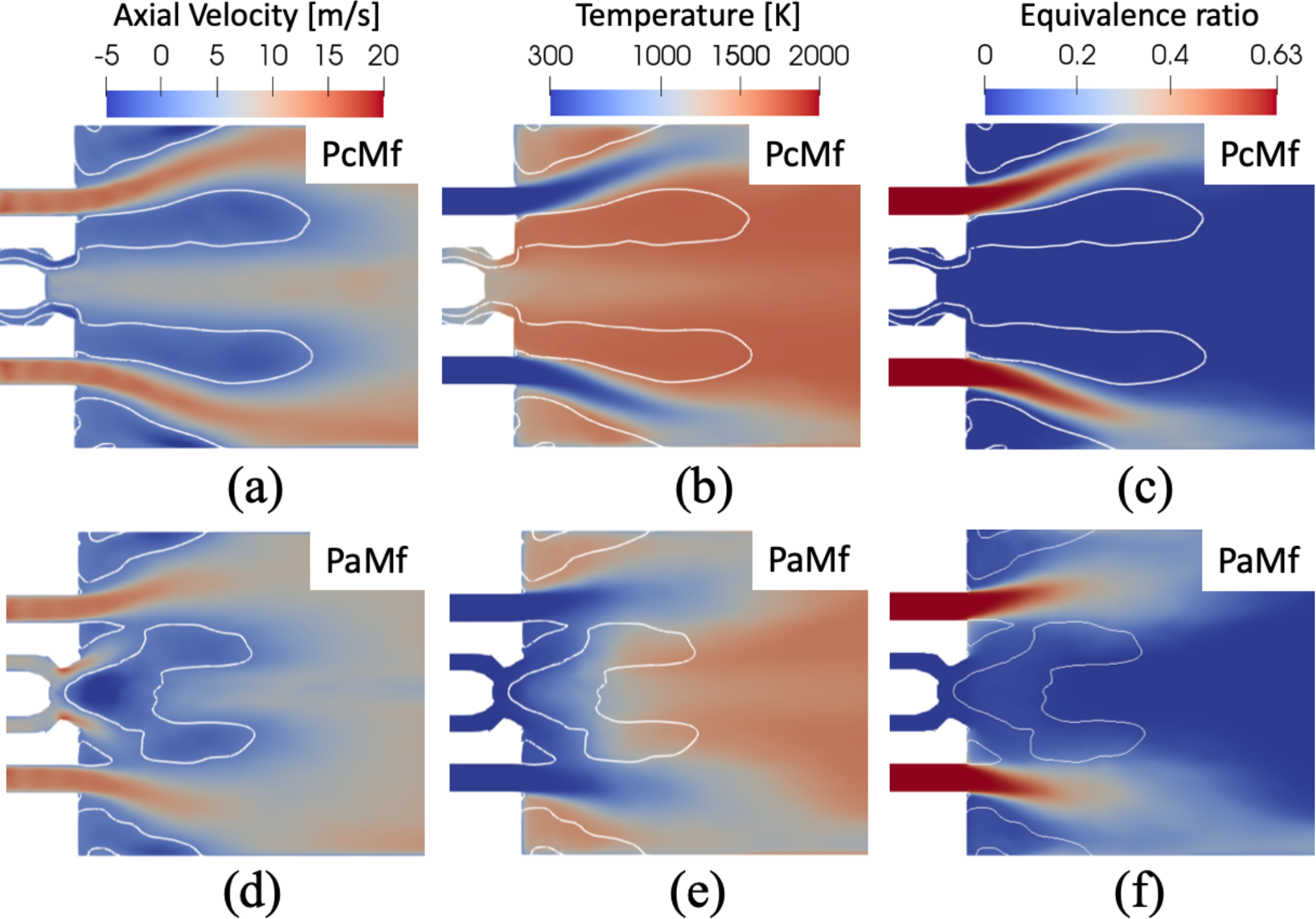}
\caption{\textcolor{red}{Time-averaged LES results for Cases $P_cM_f$ and $P_aM_f$. The distributions of axial velocity (a, d), temperature (b, e), and equivalence ratio (c, f) are shown respectively. White lines mark the zero contour of axial velocity.}}
\label{les_main_2} 
\end{figure}

\textcolor{red}{More insights on these two cases are provided by the numerical analysis in Fig.~\ref{les_main_2}. The time-averaged axial velocity distribution (Fig.~\ref{les_main_2}(a))} sees a pair of primary recirculation zones (PRZs) resembling a ``double kidney'' shape. This is because as the pilot stage is closed, the recirculation zones are widely separated. A pair of corner recirculation zones (CRZs) are also found. Due to the recirculation of hot reaction products, both the PRZs and CRZs remain high temperature as shown in Fig.~\ref{les_main_2}(b), enabling the stabilisation of the V-shape flame. The flame in Case $P_aM_f$ is instead lifted-off and tackles the side wall. In this configuration, the pilot is supplied with pure air leading to a different PRZ structure (Fig.~\ref{les_main_2}(d)) with a horseshoe shape. Due to the cold stream from the pilot stage, the temperature in the PRZ is low, as shown in Fig.~\ref{les_main_2}(e). This low-temperature PRZ prevents combustion and pushes the flame downstream, resulting in a lifted flame as shown in Fig.~\ref{les_main_1}(c). 

\begin{figure}[htbp]
\centering
     \subfigure[\label{spectrum-m1}]{%
       \includegraphics[width=0.48\textwidth]{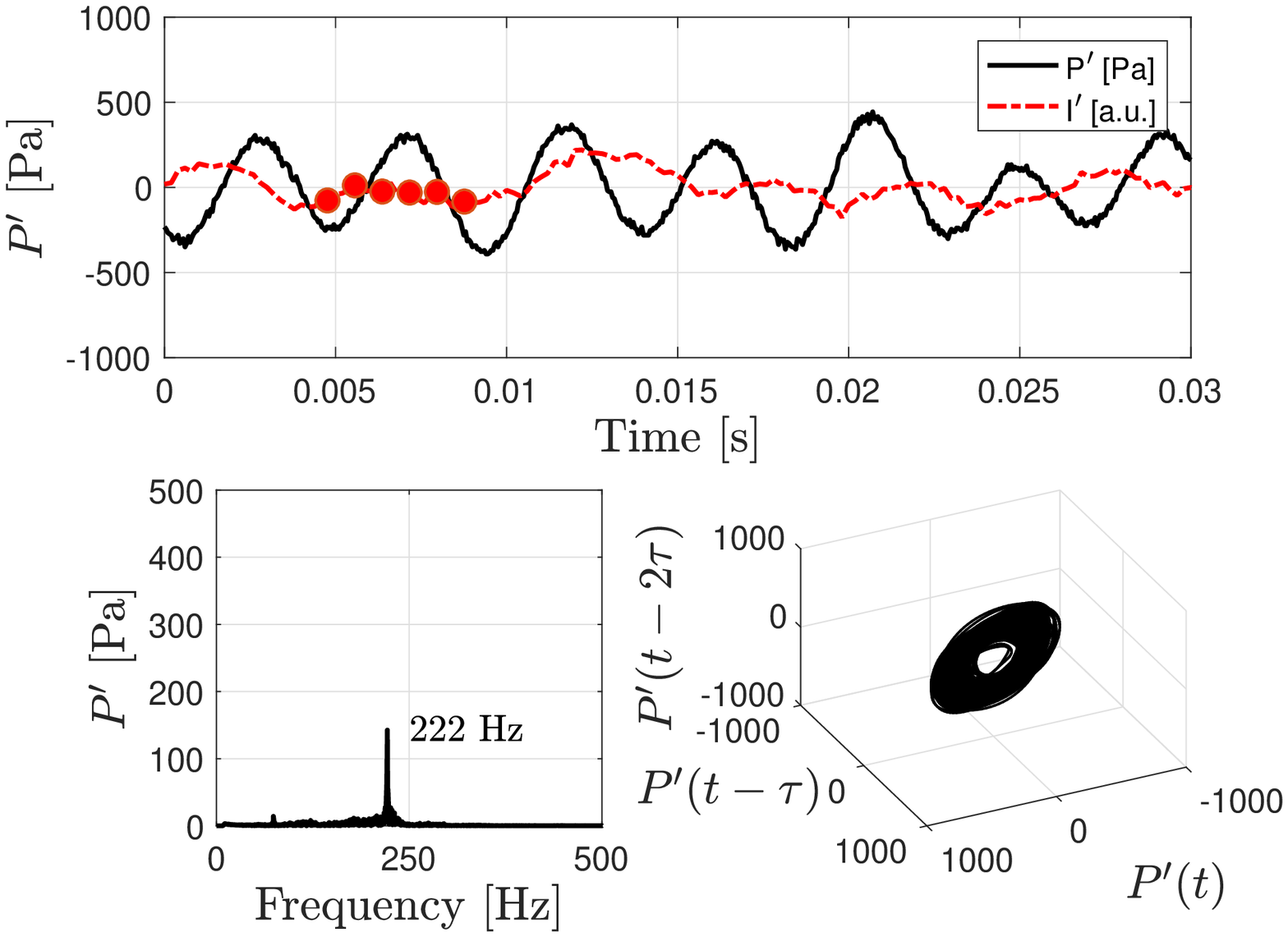}
     }
     \subfigure[\label{spectrum-m2}]{%
       \includegraphics[width=0.48\textwidth]{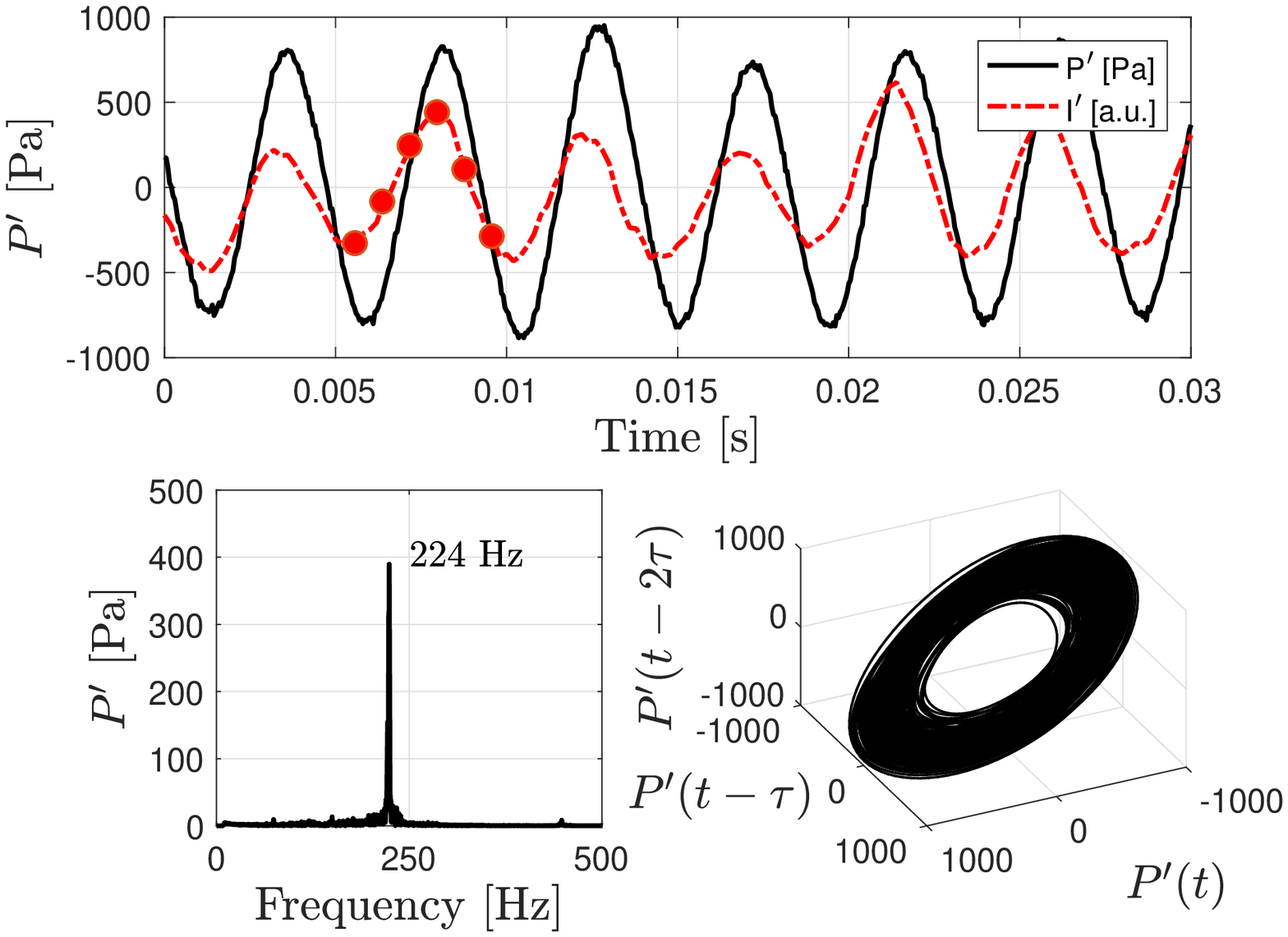}
     }\\
     \subfigure[\label{flame-m1}]{%
       \includegraphics[height=0.2\textwidth]{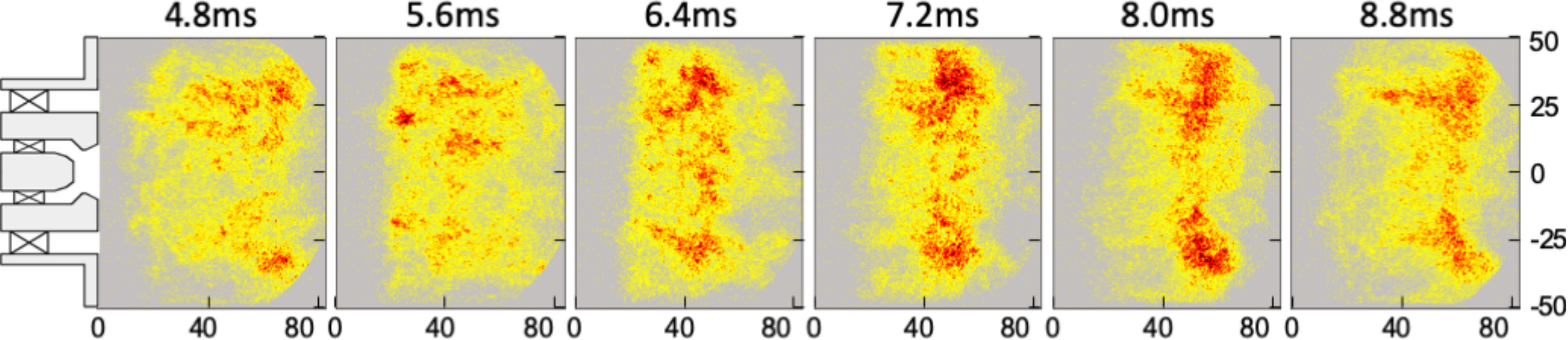}
     }\\
     \subfigure[\label{flame-m2}]{%
       \includegraphics[height=0.2\textwidth]{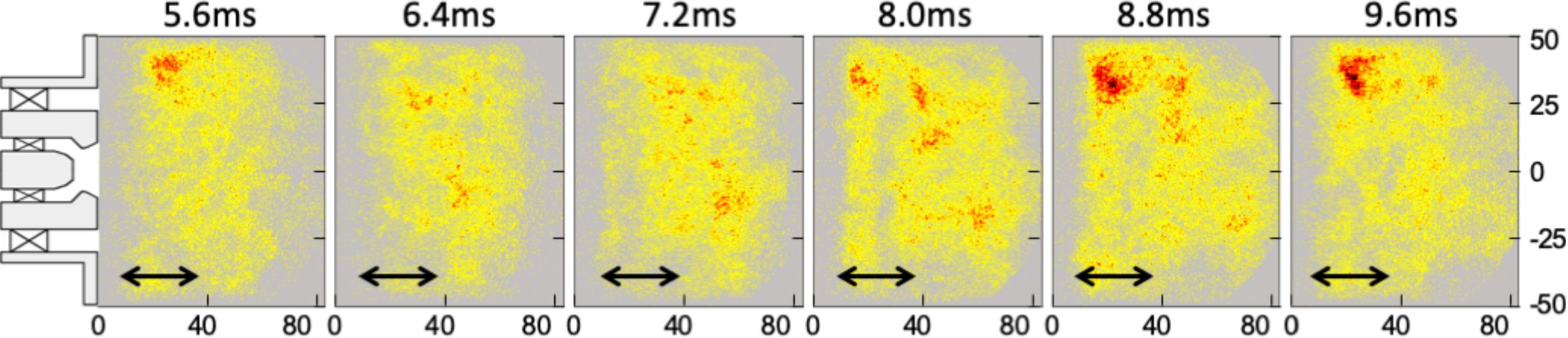}
     }\\
\caption{Pressure fluctuations and flame dynamics of the main flame mode. Comparison of pressure fluctuations of the S3 sensor for Cases $P_cM_f$ (a) and $P_aM_f$ (b). The time-series are shown at the top, while the spectra and the phase space trajectories shown at the bottom left and right, respectively. Sequential flame images of Cases $P_cM_f$ are shown in (c) and $P_aM_f$ in (d) (marked as red dots in the time-series plots).}
\label{spectrum-main} 
\end{figure}

The pressure spectra for Cases $P_cM_f$ and $P_aM_f$ of S3 sensor are shown in Fig.~\ref{spectrum-main}(a, b). Both cases are unstable with oscillations at almost the same frequency, i.e., 222 Hz for $P_cM_f$ and 224 Hz for $P_aM_f$. However, the peak amplitude in the spectrum of Case $P_aM_f$ is about 400 Pa. This agrees with what is shown in the time series plots. The phase difference between $P'$ and $I'$ in Case $P_aM_f$(Fig.~\ref{spectrum-m2}) is about 38$^\circ$. In this configuration the Rayleigh criterion~\cite{rayleigh1878explanation} is satisfied and the coupling between $P'$ and $I'$ is enhanced. Contrarily, in Fig.~\ref{spectrum-m1}, the phase-shift between $P'$ and $I'$ is larger, thereby obstructing the amplification of the oscillations.

\begin{figure}[htbp]
\centering
     \subfigure[\label{dmd-m1}]{%
       \includegraphics[width=1\textwidth]{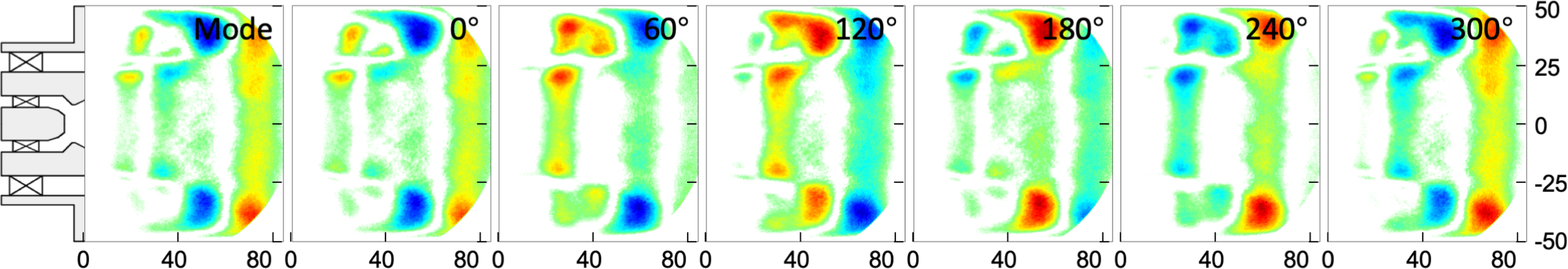}
     }\\
     \subfigure[\label{dmd-m2}]{%
       \includegraphics[width=1\textwidth]{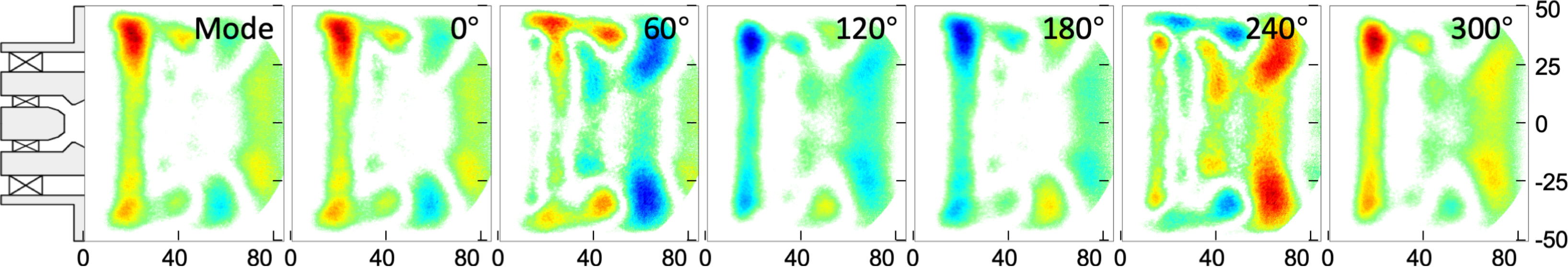}
     }\\
          \subfigure[\label{rayleigh_m1}]{%
       \includegraphics[width=0.3\textwidth]{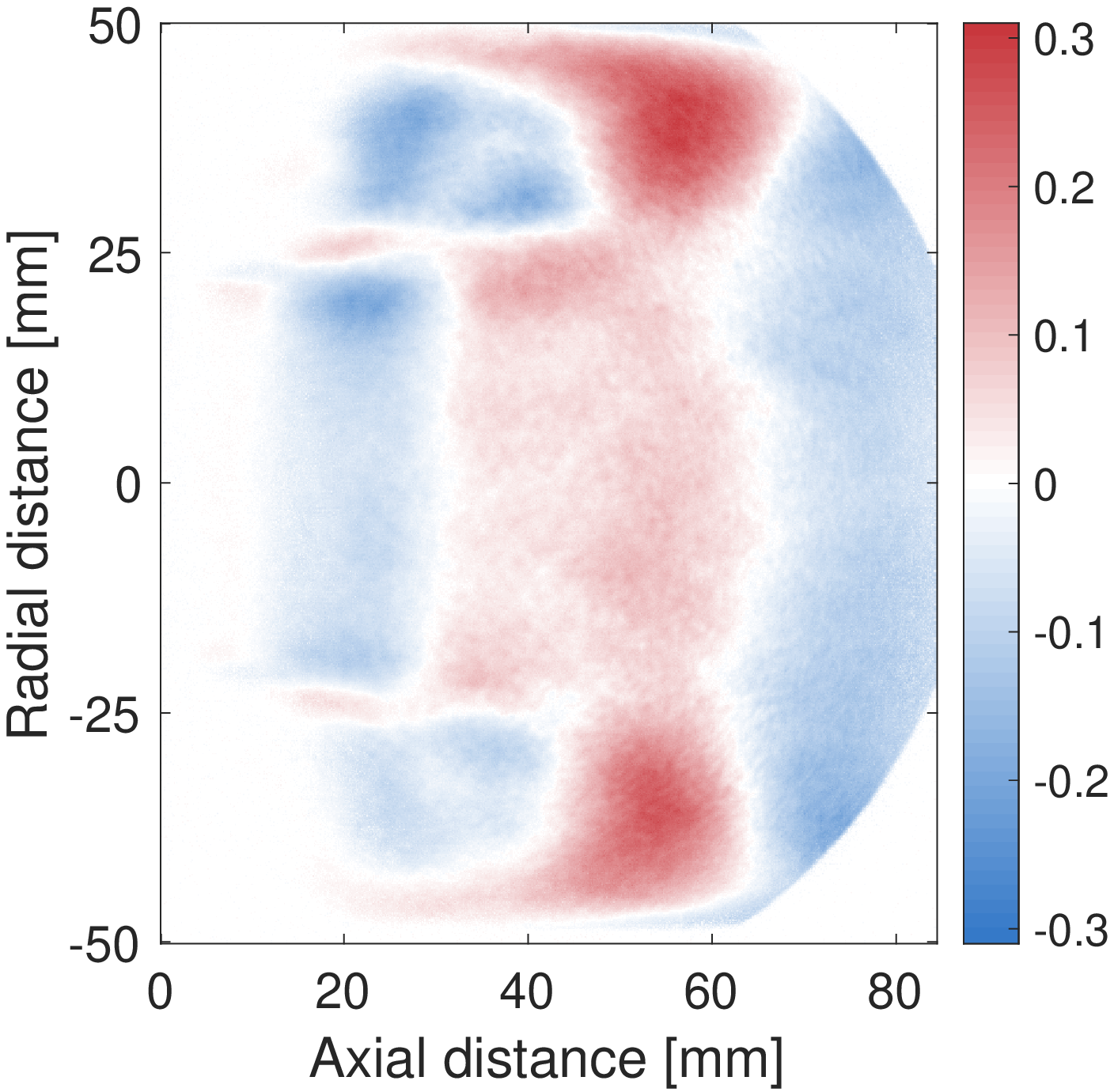}
     }
     \subfigure[\label{rayleigh_m2}]{%
       \includegraphics[width=0.3\textwidth]{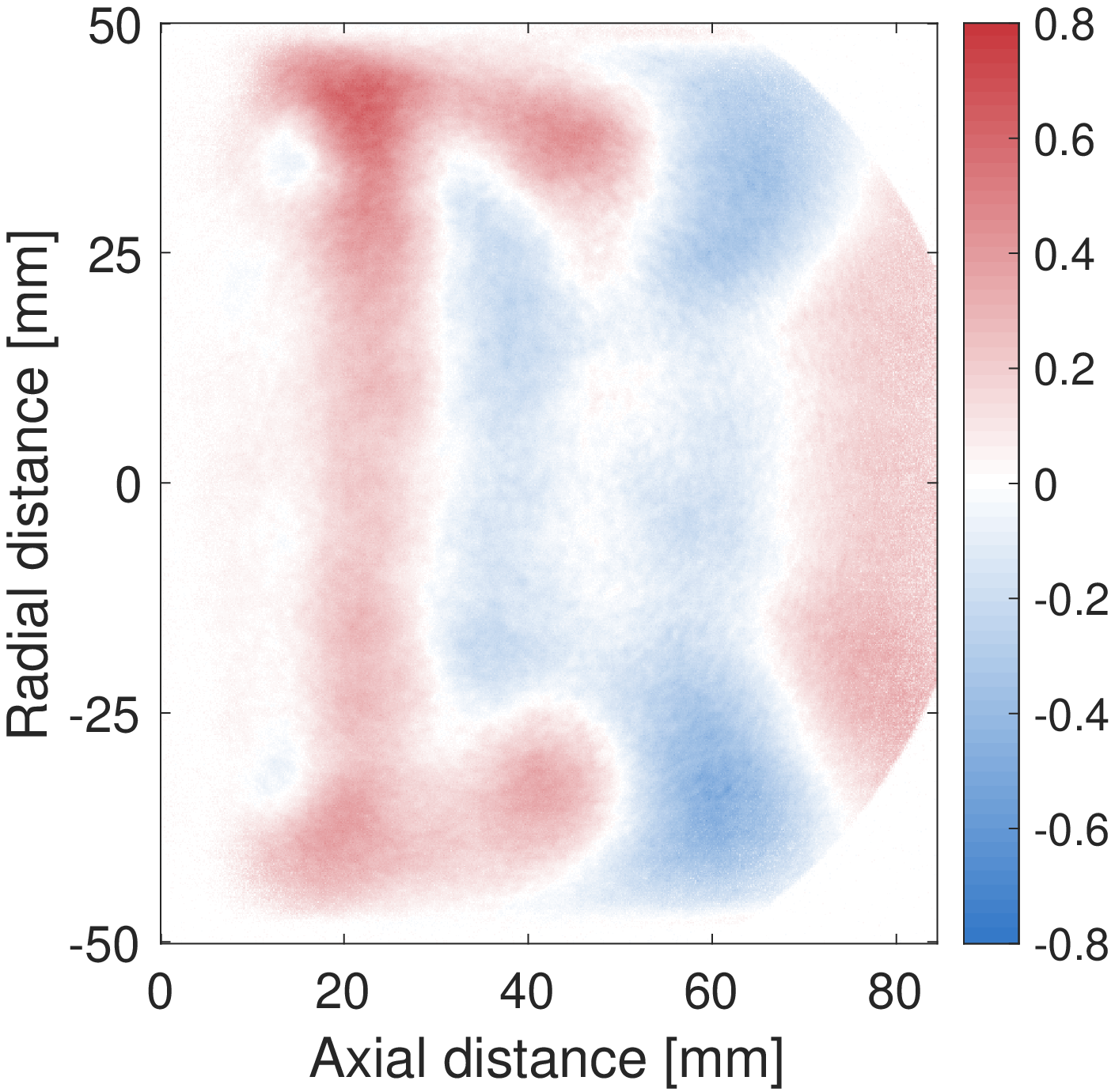}
     }\\
\caption{DMD modes (a, b) and local Rayleigh index maps (c, d) for Cases $P_cM_f$ and $P_aM_f$, respectively.}
\label{flame_m} 
\end{figure}

The sequential flame images of Cases $P_cM_f$ and $P_aM_f$ are shown in Fig.~\ref{spectrum-main}(c, d). Note that the flame in Case $P_cM_f$ is slightly lifted off from the burner and the flame surface is distorted subject to acoustic perturbations. It is also seen that the flame intensity in Fig.~\ref{flame-m1} does not vary much compared to Fig.~\ref{flame-m2}. For Case $P_aM_f$, the flame is lifted off, so the flame surface is more distributed and no clear edge can be seen in Fig.~\ref{flame-m2}. However, fluctuation of flame intensity can still be identified in the lift-off region as marked by double-end arrows. This will be more clearly seen from the DMD results in Fig.~\ref{dmd-m2}.

The DMD mode of Case $P_cM_f$ is shown in Fig.\ref{dmd-m1}. Contrary to Case $P_fM_a$, the convective motion of flame dynamics is clearly seen in the phase sequences. The coherent structure of flame moves downstream step by step. Note that no air is supplied to the pilot stage in Case $P_cM_f$, therefore the methane in the main stage is not diluted and is kept fully premixed. As for Case $P_aM_f$, DMD modes in Fig.~\ref{dmd-m2} do not exhibit convective motion but feature alternative red-blue bulk oscillation in the lift-off region.

The Rayleigh index maps of Cases $P_cM_f$ and $P_aM_f$ are shown in Fig.~\ref{flame_m}(c, d). As the driving is much stronger than damping, their respective locations with respect to the flame motion could be more interesting. In Fig.~\ref{rayleigh_m1}, the damping zone (blue) is found in the upstream of the flame, corresponding to the CRZ and PRZ, while the driving zone (red) concentrates in the flame-wall-interaction region. This flame-wall-interaction was found in this rig before and discussed in~\cite{han2019flame}. For Case $P_aM_f$, as the flame is lifted off, the main driving zone is located in the lift-off region while the convective delay gives a damping zone further downstream near the side wall. In this case, the flame is lifted with high turbulence and can only stabilise along the side wall.

\subsection{Stratified flame mode}
 \label{Stratified flame mode}

\begin{figure}[htbp]
\centering
\includegraphics[width=0.4\textwidth]{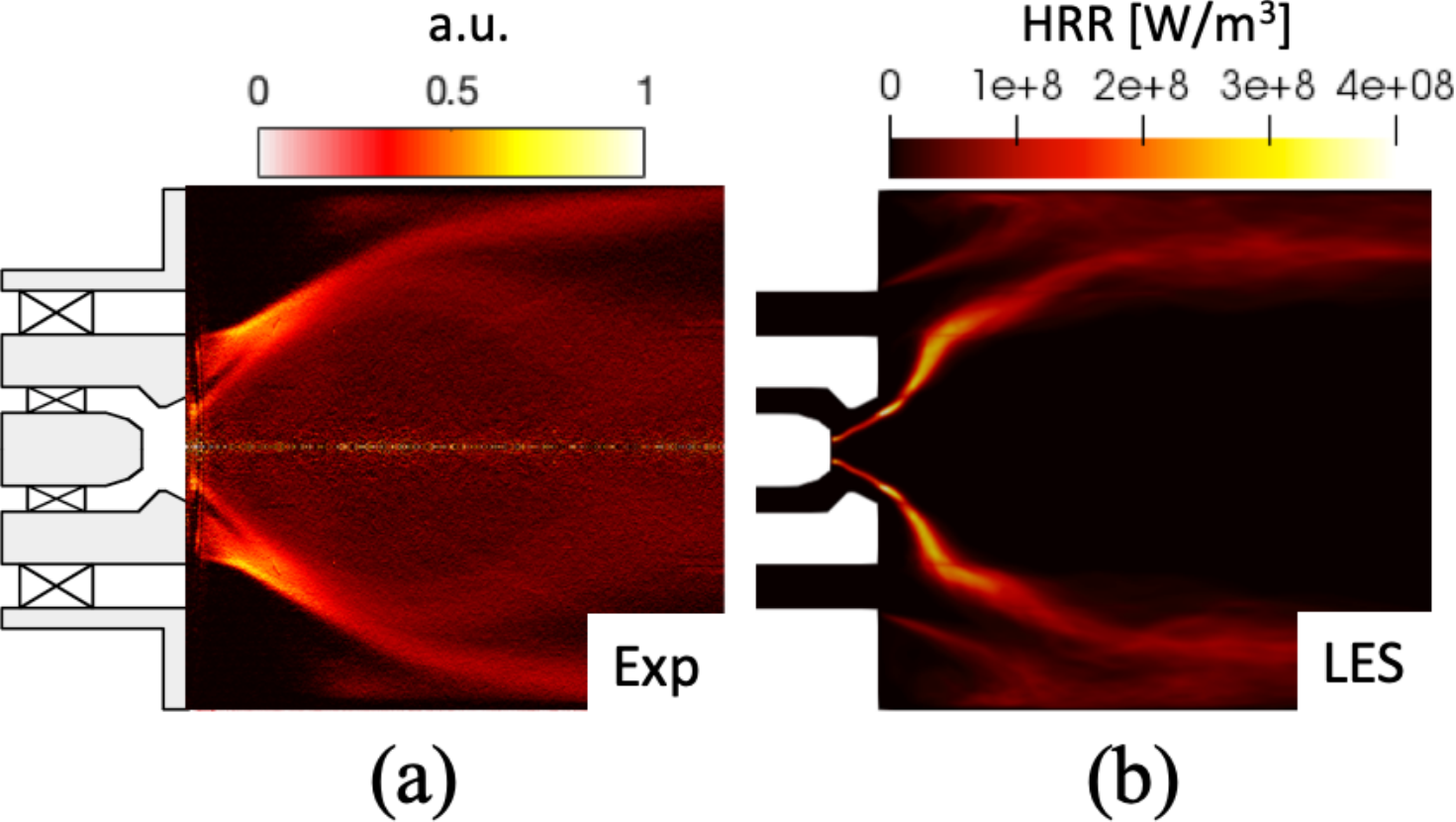}
\caption{\textcolor{red}{Time-averaged flame shapes captured by the DSLR camera (a) and LES HRR (b) for Case $P_fM_f$.}}
\label{les_strat_1} 
\end{figure}

In Case $P_fM_f$, premixed fuel-air mixtures are injected from both the pilot and main stages with different equivalence ratios to give a stratified swirl flame, similar to that widely used in LPP combustors. The \textcolor{red}{measured} time-averaged flame image is shown in \textcolor{red}{Fig.~\ref{les_strat_1}(a)}. The pilot flame is anchored inside the pilot stage and develops along the CSL while the main flame is anchored at the exit of the main stage along the ISL. The two flame sheets merge downstream and then impinge towards the side wall. The overall flame shape is well captured by LES \textcolor{red}{reported in} Fig.~\ref{les_strat_1}(b), with some mismatch at the root of the main flame.

\begin{figure}[htbp]
\centering
\includegraphics[width=0.6\textwidth]{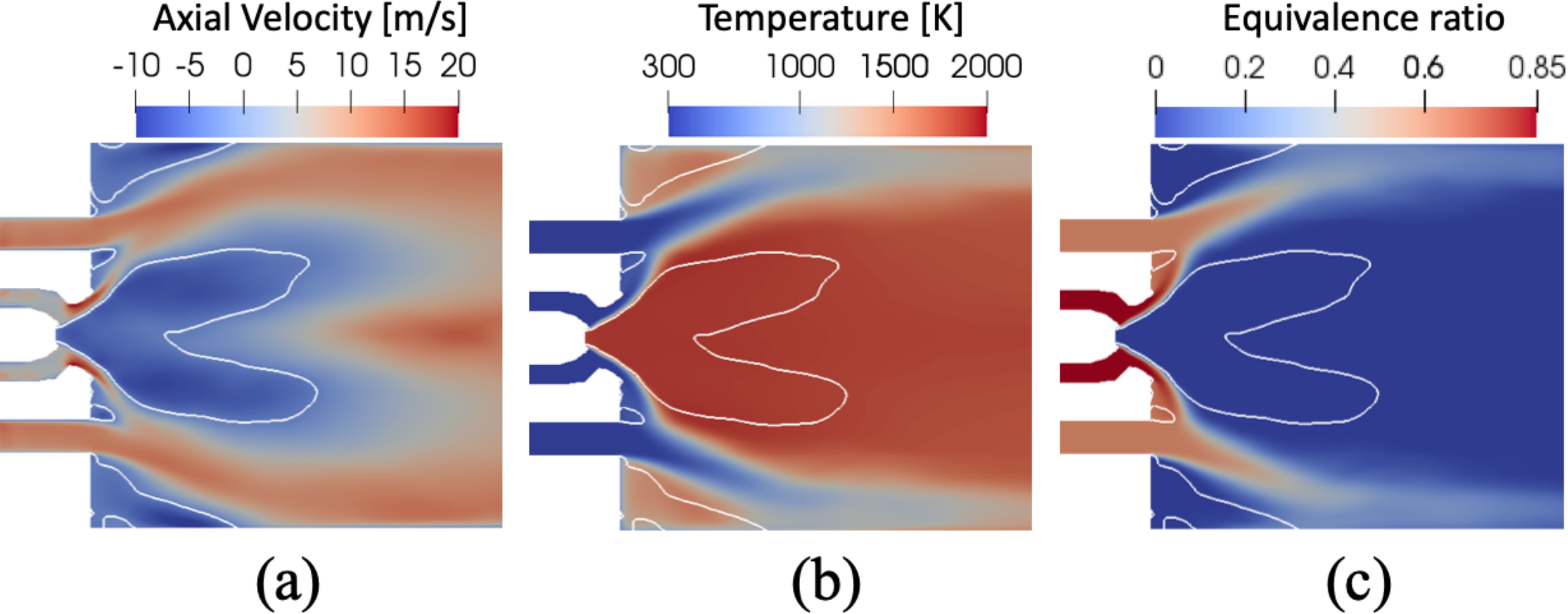}
\caption{\textcolor{red}{Time-averaged LES results for Case $P_fM_f$. The distributions of axial velocity (a), temperature (b), and equivalence ratio (c) are shown respectively. White lines mark the zero contour of axial velocity.}}
\label{les_strat_2} 
\end{figure}

\textcolor{red}{With respect to Case $P_aM_f$ (Fig.~\ref{les_main_2}(d)), a more extended horseshoe shape PRZ is revealed by LES for this case (Fig.~\ref{les_strat_2}(a)).} As the pilot stage is supplied with the methane-air mixture, the PRZ is now full of high-temperature gases, providing a suitable condition for flame stabilization. The merging and mixing of the pilot and main streams can also be seen in Fig.~\ref{les_strat_2}(a), leading to an equivalence ratio gradient at the merging region as shown in Fig.~\ref{les_strat_2}(c). \textcolor{red}{As expected, the PRZ is filled with high temperature gas, which helps the stabilization of the stratified flame (Fig.~\ref{les_strat_2}(b)).}

\begin{figure}[htbp]
\centering
     \subfigure[\label{spectrum-s}]{%
       \includegraphics[width=1\textwidth]{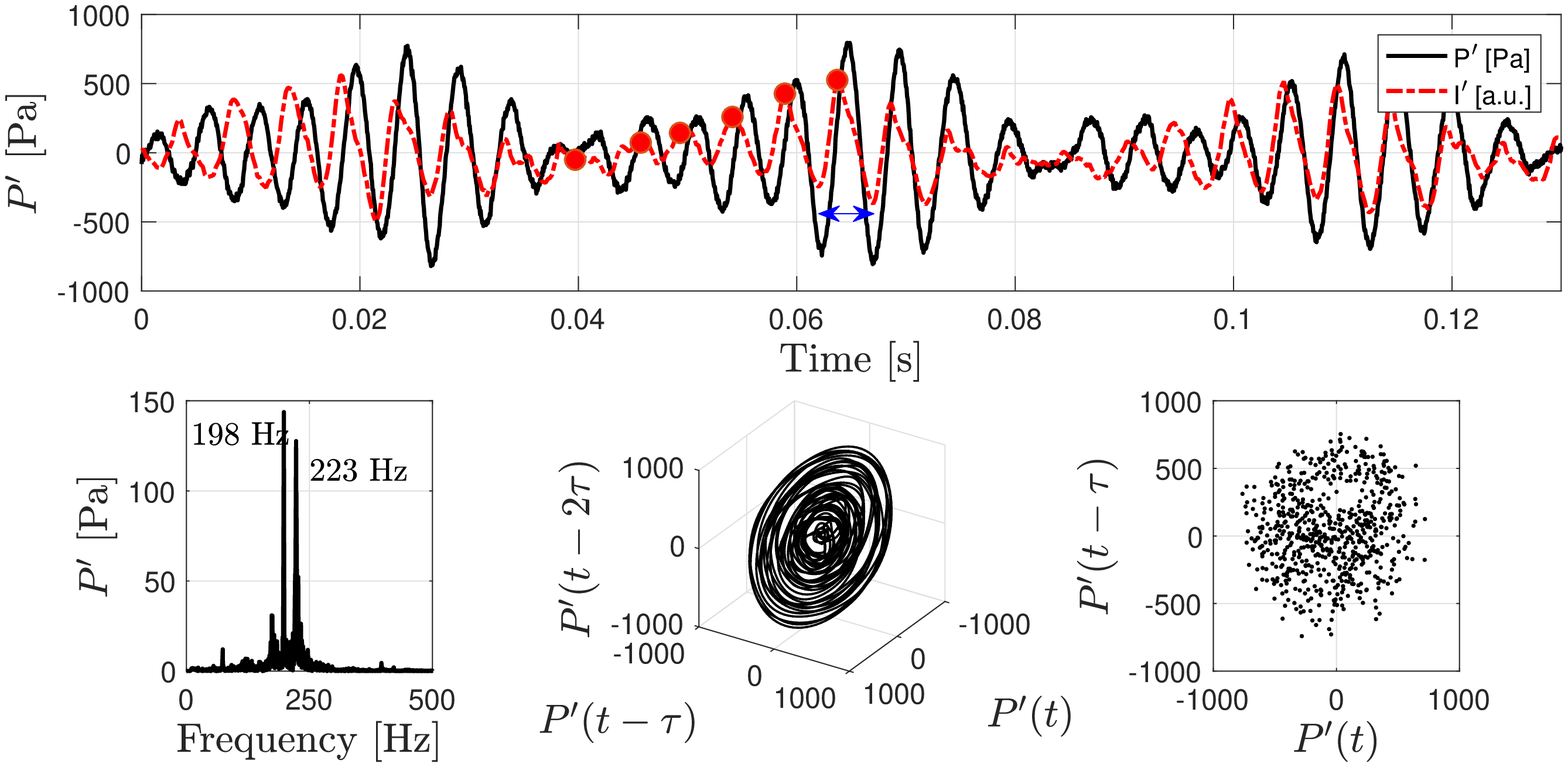}
     }
     \subfigure[\label{flame-s-long}]{%
       \includegraphics[height=0.2\textwidth]{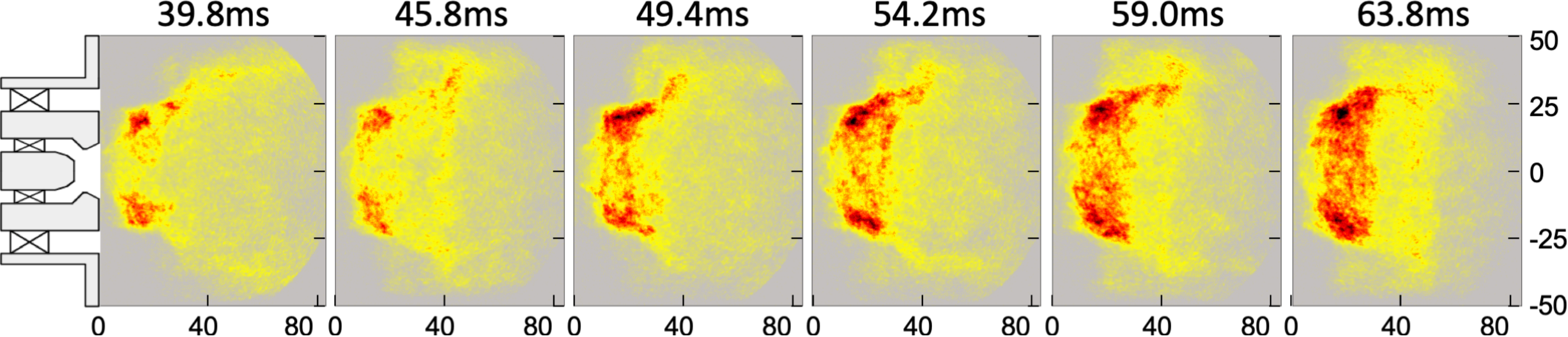}
     }\\
     \subfigure[\label{flame-s-short}]{%
       \includegraphics[height=0.2\textwidth]{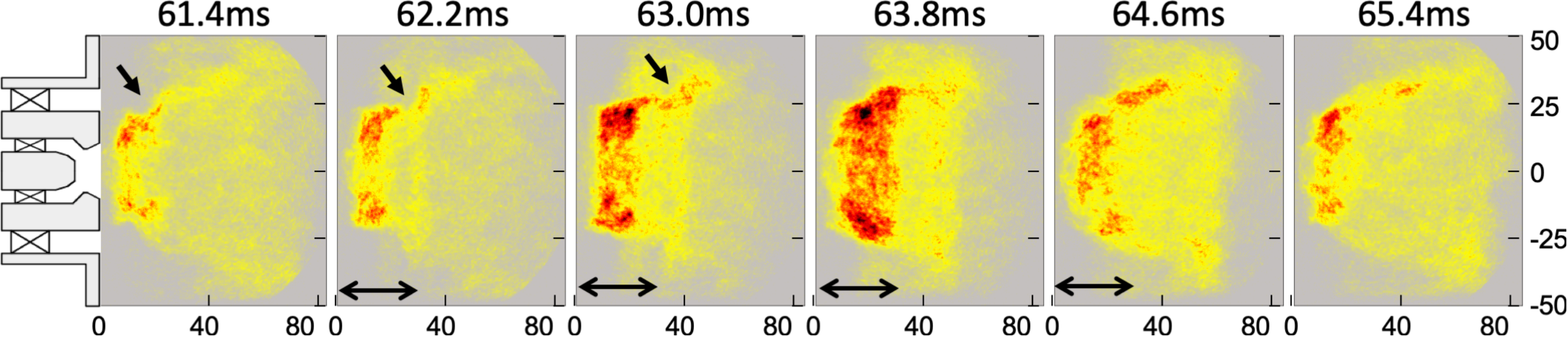}
     }\\
\caption{Beating oscillations found in Case $P_fM_f$. (a) The fluctuations of pressure  and flame intensity. Time-series plot is shown on the top, while the spectrum (left), the phase space trajectory (middle), and the Poincaré map (right) are shown in the bottom. (b) Sequential flame images along the envelope line (marked as red dots). (c) Sequential flame images during one period (marked as the blue arrow).}
\label{46402} 
\end{figure}

Compared to previous cases, the present stratified flame case sees a very different combustion instability oscillation. Figure~\ref{spectrum-s} reports the recorded pressure oscillations in which $P'$ is seen to be periodic but shows a low-frequency envelope in its oscillation amplitude. In the spectrum, this corresponds to the presence of two peaks with close frequency: the first peak is 198 Hz and has an amplitude of 145 Pa, and the second is 223 Hz with a similar amplitude.

In a dynamic system, this is usually referred to as ``beating''~\cite{kim2019experimental,weng2016investigation}. As shown in Eq.~\ref{eq_beating}, it includes two sinusoidal signals with close frequencies. In the current case, $f_1=198$ Hz, $f_2=223$ Hz, and $f_{1-2}$=25 Hz -- this frequency difference being consistent with the frequency of the signal envelope shown in Fig.~\ref{46402}. In the time domain signal, the heat release perturbation, $I'$, slightly lags the pressure perturbation, $P'$, and also exhibits beating oscillations. Phase space reconstruction shows that the oscillation is not a limit cycle. The Poincaré map is also plotted on the right side, which is the intersection of the phase space trajectory with a certain lower-dimensional subspace and is widely used in nonlinear time series analyses~\cite{kantz2004nonlinear}. In a Poincaré map, a pure sinusoidal signal is projected onto a discrete point, while a quasi-periodic signal into a closed ring. In the observed beating oscillations, the Poincaré points are scattering along a ring, indicating the signal is a quasi-periodic oscillation accompanied with strong noise~\cite{kabiraj2015chaos}.

To evaluate the strength of beating oscillations, a beating index, B, is defined as follow:
\begin{equation}
B= \frac{|A_1| \times |A_2|}{max(|A_1|, |A_2|)^2},
\label{eq_index}
\end{equation}
where $A_1$ and $A_2$ are Fourier amplitudes of the two sinusoidal signals. The results in Fig.~\ref{46402} were obtained by considering equivalence ratios $\phi_p$=0.85 and $\phi_m$=0.63. Extra experiments with different $\phi_p$ and $\phi_m$ have also been conducted. The beating index results for these different cases are summarised in Table~\ref{tab_beating}. One can see that beating oscillations are only found over a narrow window of operating conditions and the beating index reaches its maximum at $\phi_p$=0.85 and $\phi_m$=0.63. This indicates that beating oscillations are highly sensitive to the equivalence ratio variations. Note that when the main and pilot flames are independently operated (see Case $P_fM_a$ in which $\phi_p$=0.85 is studied and Case $P_aM_f$ in which $\phi_m$=0.63 is studied), intense thermoacoustic instabilities have been observed. This indicates that beating oscillations in Case $P_fM_f$ may originate from the interactions between the pilot and main flame dynamics. The frequencies of the beating cases in Table~\ref{tab_beating} are shown in Table A in Supplemental Material.

\begin{table}[htbp]
\caption{Dependence of beating index on equivalence ratios of the pilot and main flames.}
\begin{center}
\label{tab_beating}
\begin{tabular}{c|cccc}
\hline
\diagbox{$\phi_p$}{$\phi_m$} & 0.58  & 0.61  & 0.63  & 0.65  \\\hline
0.48     & 0 & 0 & 0 & 0.27 \\
0.67     & 0.08 & 0.12 & 0.16 & 0.17 \\
0.85     & 0 & 0.27 & \textbf{0.86} & 0.79 \\
1.04     & 0 & 0 & 0 & 0 \\
1.22     & 0 & 0 & 0 & 0\\\hline
\end{tabular}
\end{center}
\end{table}

We now study Case $P_fM_f$ in more detail. Two sequences of flame images are shown in Fig.~\ref{46402}(b, c). The first is chosen along the beating envelope from 39.8 ms to 63.8 ms, as shown by red dots in Fig.~\ref{46402}(a). It can be seen that the overall flame shape remains similar with the main heat release region locate near the merging region between the pilot and main flames, but the flame intensity increases gradually since the perturbation amplitude increases. In one oscillation cycle, between 61.4 ms and 65.4 ms as shown by the blue arrow in Fig.~\ref{46402}(a), detailed flame dynamics are shown in sequence in Fig.~\ref{flame-s-short}. Two types of flame dynamics are identified. From 61.4 ms to 63.0 ms, flame surface wrinkling can be clearly seen by following the black arrow marks, which is most probably caused by the velocity perturbation and vortex roll-up effect~\cite{palies2010combined}; this being a typical flame dynamics of premixed swirl flame~\cite{candel2014dynamics}. Significant flame intensity fluctuation at the merging region between the pilot and main flames can also be clearly seen, e.g. between 62.2 ms and 64.6 ms by following the double-end arrows. This periodic bulk fluctuation is similar to what has been observed in Fig.~\ref{flame-p2} for $P_fM_a$. As stated in Section~\ref{Pilot flame mode}, this bulk flame motion has similar characteristics of the typical dynamics of partially premixed flame~\cite{stohr2017interaction,wang2019combustion}. Note again that in this merging region, equivalence ratio fluctuation could exist due to the unsteady mixing of the two different premixed streams.

\begin{figure}[htbp]
\centering
     \subfigure[\label{phase_strat_198}]{%
       \includegraphics[width=1\textwidth]{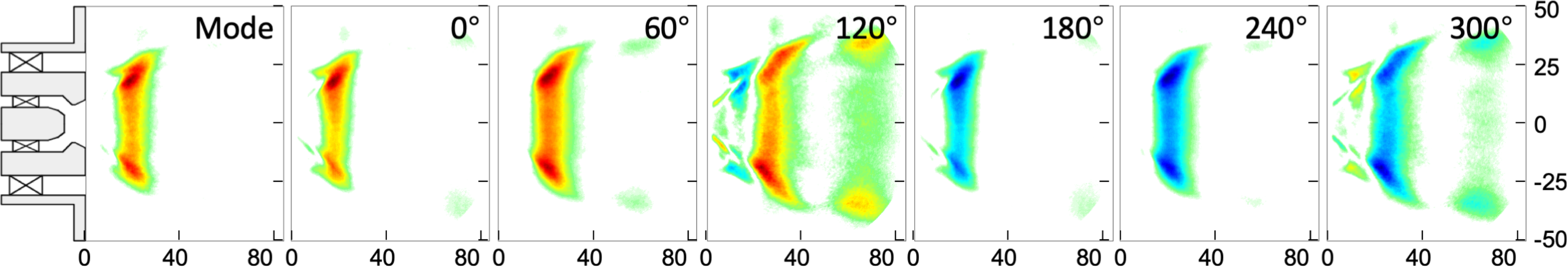}
     }\\
     \subfigure[\label{phase_strat_223}]{%
       \includegraphics[width=1\textwidth]{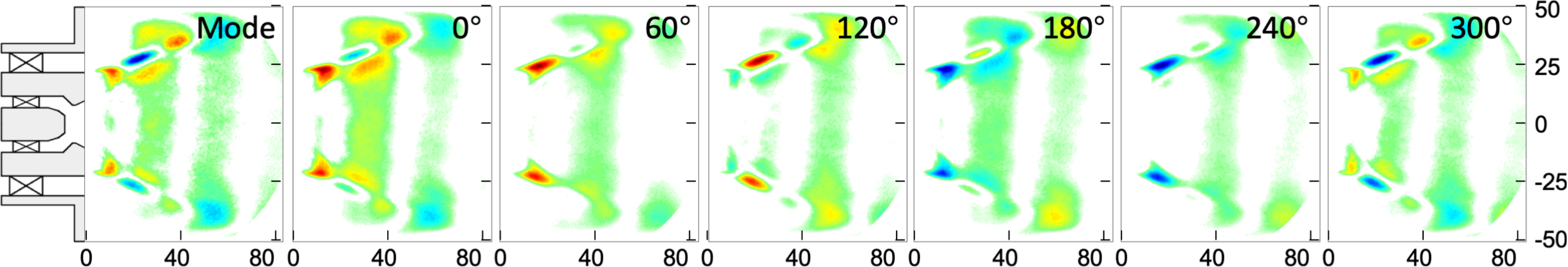}
     }\\
     \subfigure[\label{RI_all}]{%
       \includegraphics[width=0.3\textwidth]{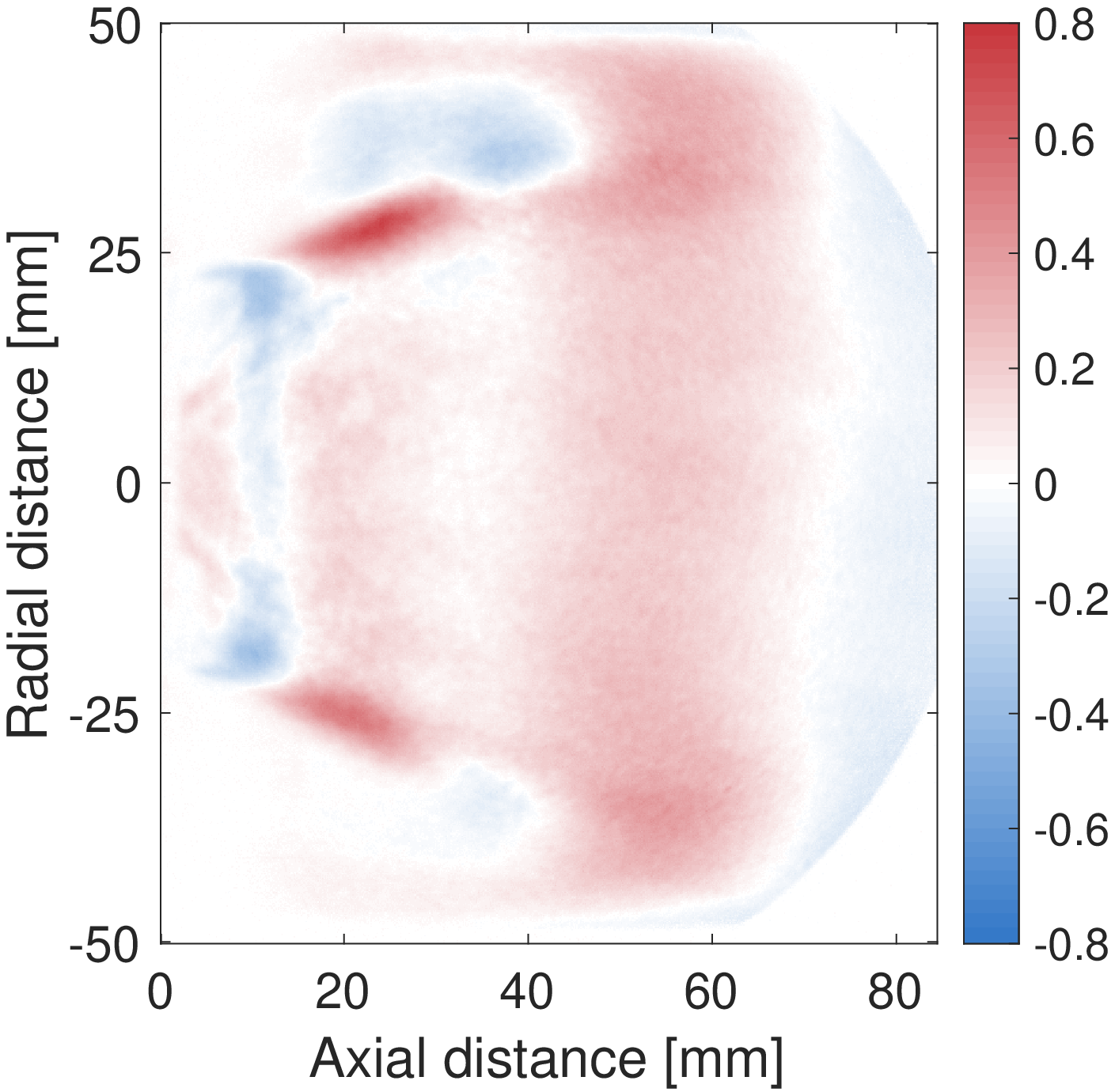}
     }
     \subfigure[\label{RI_198}]{%
       \includegraphics[width=0.3\textwidth]{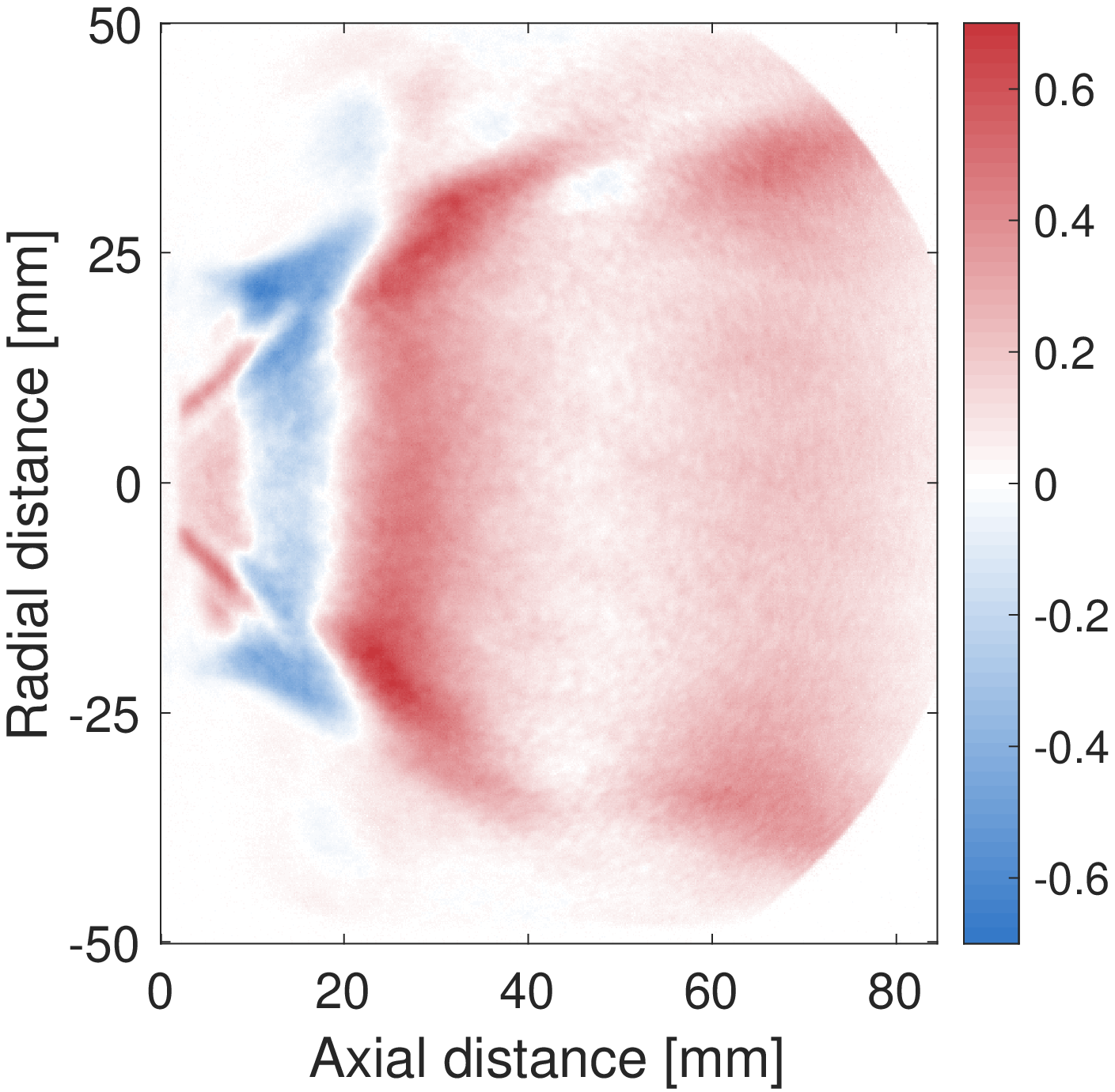}
     }
     \subfigure[\label{RI_223}]{%
       \includegraphics[width=0.3\textwidth]{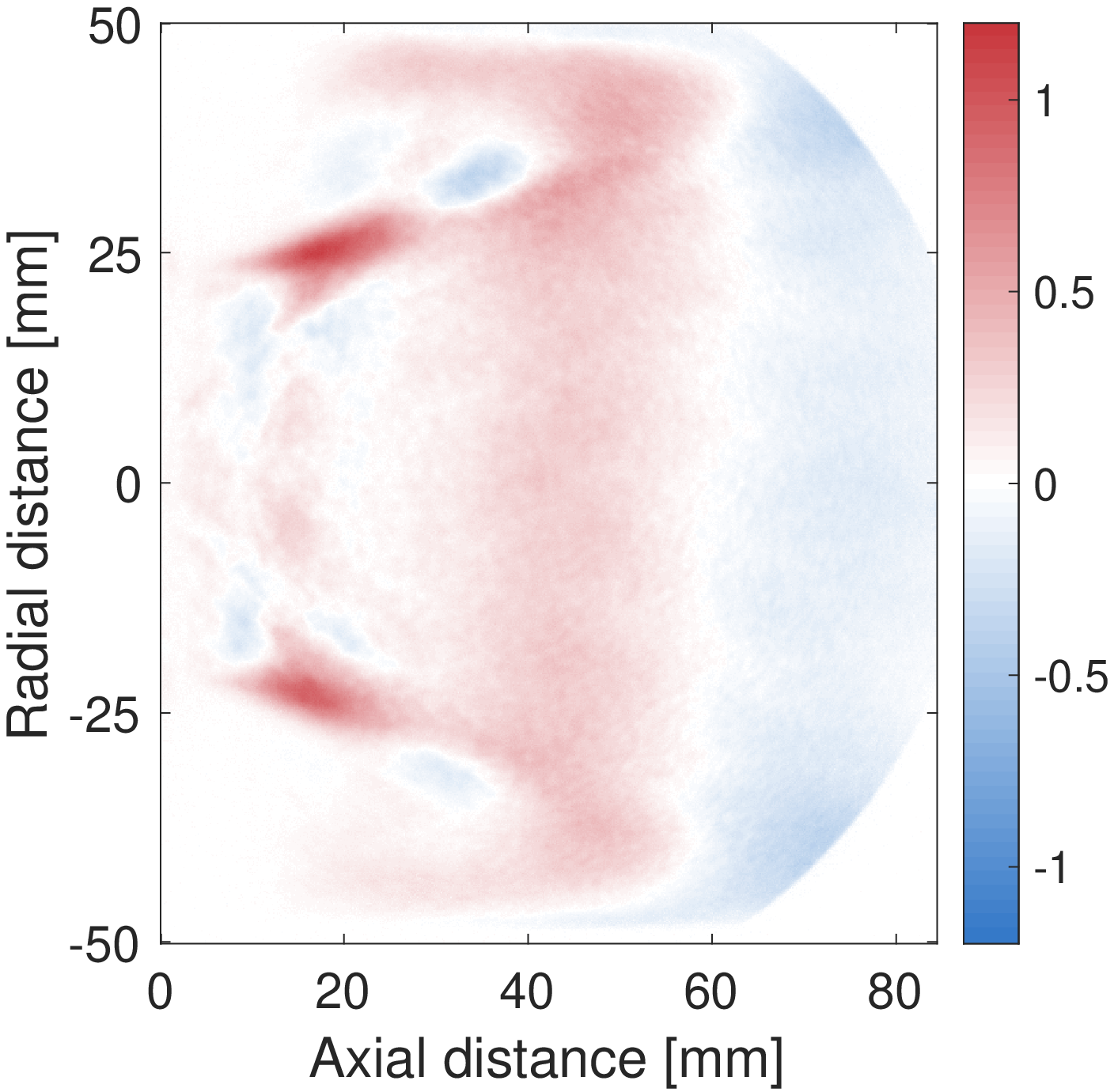}
     }\\
\caption{DMD modes (a, b) and local Rayleigh index maps (c-d) of Case $P_fM_f$. Dynamics of DMD modes at 198 Hz (a) and 223 Hz (b). Rayleigh index map via the original data (c), Rayleigh index map at 198 Hz (d) and 223 Hz (e).}
\label{flame_S} 
\end{figure}

In order to see the flame dynamics more clearly, DMD modes of the flame images are extracted for each of the two peak frequencies in the oscillation spectrum, with results shown in Fig.~\ref{flame_S}(a, b). In the phase sequence of Fig.~\ref{phase_strat_198}, both the red and blue regions have nearly fixed axial locations -- proving that bulk oscillation is the dominant mechanism of flame dynamics for the 198 Hz mode. This resembles a similar bulk oscillation as shown in Fig.~\ref{dmd-p2} (of Case $P_fM_a$), indicating that both of them are most probably relevant to equivalence ratio fluctuations.

On the contrary, convective, and thus spatially distributed, coherent structures along the ISL are found at the DMD results at 223 Hz in Fig.~\ref{phase_strat_223}. It can be seen from the phase sequence of Fig.~\ref{phase_strat_223} that two red spots (one at the top and one at the bottom) start from the LRZ at 0$^\circ$, move towards the merging region at 60$^\circ$ with an increased amplitude. The two red spots move further downstream along the ISL at 120$^\circ$ and gradually disappear at 180$^\circ$. Two new blue spots are then generated from the LRZ at 180$^\circ$ and move downstream in a similar way. This convective flame motion has similarities to the one shown in Fig.~\ref{dmd-m1} (of {Case $P_cM_f$}) which is obtained when only the main stage has an attached flame. Note again that similar flame dynamics have been widely seen in premixed swirl flames~\cite{candel2014dynamics}.

It can be seen that flame dynamics in Case $P_fM_f$ have two types of oscillations which respectively share similarities with the ones observed when the main and pilot flames are independently operated. This reveals the complexity of the current stratified flame, and also indicates the fundamental mechanisms for the beating phenomena. In order to see how the two types of flame dynamics are correlated to the thermoacoustic oscillations, three local Rayleigh index maps are shown in Fig.~\ref{flame_S}(c-e). The Rayleigh index map using the original data for both the pressure and flame images is plotted in Fig.~\ref{RI_all}. Similar to Section 3.2, we pay more attention to the respective locations of the damping and driving zones. Damping zones (blue) are identified both near the pilot-main flame merging region and outside of the flame sheet while driving zones (red) are found along with the flame sheet, near the flame merge and the flame-wall-interaction regions. We then use FFT to extract the two frequencies components for each pixel of the flame images and combine each of them with its corresponding frequency component in the pressure signal to give Local Rayleigh index for the two frequencies separately. These are shown in Fig.~\ref{flame_S}(d, e). It can be clearly seen that although a damping region is seen at the merging region of the pilot-main flame at 198 Hz, the acoustics at both frequencies are in general driven by their coupling with the corresponding flame dynamic modes. These decoupled Rayleigh index maps provide important information on how both frequencies in the beating oscillations are sustained by the complex pilot-main flame interactions.

\subsection{Acoustic analysis of beating oscillations}\label{sub:acoustics}

Based on the LES results, the change in vortex structures and related velocity fluctuations are found to be unrelated to the frequencies of the thermoacoustic oscillations. Please refer to Figs. D and E in Supplemental Material for more details. In order to understand the nature of the two frequencies, $f_1$ and $f_2$, observed in Case $P_fM_f$, a simplified acoustic analysis is now conducted. 

\begin{figure}[htbp]
\centering
     \subfigure[\label{geo1}]{%
       \includegraphics[width=0.6\textwidth]{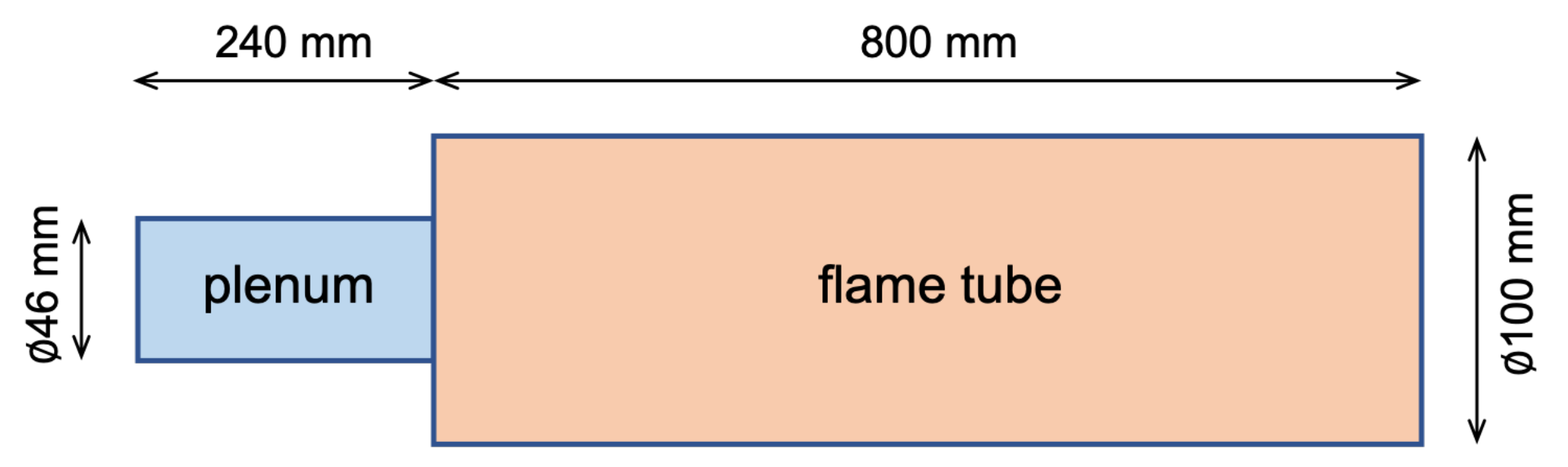}
     }\\
     \subfigure[\label{geo2}]{%
       \includegraphics[width=0.6\textwidth]{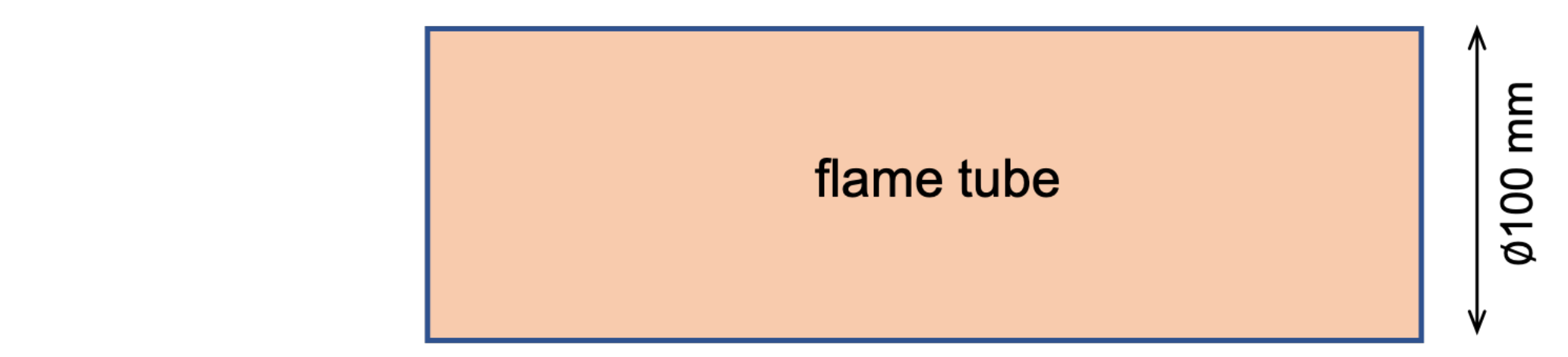}
     }\\
\caption{Simplified models for acoustic analysis: (a) \emph{Arrangement A}: plenum + flame tube, and (b) \emph{Arrangement B}: flame tube only.}
\label{geo} 
\end{figure}

As shown in Fig.~\ref{rig2}, the test rig can be considered the combination of two parts: the upstream plenum and the flame tube. Following the works of Lamraoui \emph{et al.}~\cite{amraoui2011experimental} and Schuller \emph{et al.}~\cite{schuller2012acoustic}, the presence of the sudden expansion between these two sections means the system can feature coupled or decoupled acoustic modes. The coupling index is defined as: $\Xi \simeq S_{1} / S_{2}\left(T_{u} / T_{b}\right)^{1 / 2}$, where $S_{1}$ and $S_{2}$ are the cross-sectional areas of the plenum and flame tube respectively, and $T_{u}$ and $T_{b}$ are the fresh and burnt gas temperatures respectively. The resonant frequencies can be calculated using the two simplified arrangements as shown in Fig.~\ref{geo}: 1) \emph{Arrangement A}: plenum + flame tube; 2) \emph{Arrangement B}: flame tube only. The plenum temperature $T_u$ is 300 K, while the average temperature of the flame tube, $T_b$, is obtained by averaging thermocouple measurements at different axial locations along the flame tube. Due to heat loss, $T_b$ is lower than the adiabatic temperature. $T_b$ and the coupling index $\Xi$ are listed in Table~\ref{tab_freq}. As $T_b$ is higher for Case $P_fM_f$ and $P_cM_f$ than for Case $P_fM_a$, $\Xi$ is lower for Cases $P_cM_f$ and $P_fM_f$, indicating a weaker coupling between the plenum and flame tube~\cite{schuller2012acoustic}. The frequencies of the first acoustic mode for the two arrangements are calculated using the open-source toolbox OSCILOS\footnote{www.oscilos.com}~\cite{li2015time}, which can be applied to both longitudinal~\cite{li2015time,li2017numerical} and annular configurations~\cite{yang2019systematic}. For simplicity, the mean heat release of the flame is considered via imposing a mean temperature jump across the flame, but unsteady heat release is not considered. A velocity node is imposed at the inlet and a pressure node at the outlet for both.

\begin{table}[htbp]
\caption{Predicted acoustic resonant frequencies for the test rig.}
\begin{center}
\label{tab_freq}
\begin{tabular}{cccccc}
\hline
Case     &  Averaged $T_b$  & $\Xi$  &    $f_A$  &   $f_B$ & Experimental results \\ \hline
$P_fM_a$ &    746 K  &   0.137  & 156.2 Hz     &  168.5 Hz   &     159 Hz        \\
$P_cM_f$  &    1294 K   & 0.102  & 198.5 Hz        &  218.7 Hz      &     224 Hz      \\
$P_fM_f$   &   1379 K   & 0.099  & 203.9 Hz & 225.4 Hz     &   198/223 Hz     \\ \hline
\end{tabular}
\end{center}
\end{table}

Table~\ref{tab_freq} shows the resonant frequencies of the first longitudinal mode predicted by the two models for three unstable operating conditions, i.e., Cases $P_fM_a$, $P_cM_f$ and $P_fM_f$.
The frequency in Case $P_fM_a$ is captured by \emph{Model A}. In Case $P_fM_a$, only the pilot flame is operated, so the average temperature in the flame tube appears not to be high enough to sustain a decoupled mode. Conversely, a decoupled mode does occur for the main flame mode (Case $P_cM_f$); thermoacoustic instability occurs close to a decoupled flame tube area. Finally, in Case $P_fM_f$, both of the two predicted resonant frequencies, i.e., $f_A$=203 Hz and $f_B$=225.4 Hz agree approximately with the experimental observations $f_1$=198 Hz and $f_2$=223 Hz. The frequency difference, $f_{A-B}$=20 Hz, also agrees approximately with the experimental beating oscillations frequency of $f_{1-2}$=25 Hz. In contrast to the two previous cases, this stratified flame couples simultaneously with both acoustic modes, resulting in unique beating oscillations. To further validate this simplified acoustic analysis, we have conducted an extra experiment with a longer flame tube of 1200 mm in length. The frequency of beating oscillation can also be predicted by the same approach used here. Please refer to Supplemental Material for details (\emph{cf.} Fig. F and Table B).

Based on the flame dynamics and Rayleigh Index analysis in Sections 3.1-3.3, we can conclude that the beating oscillations come from the interactions between the pilot and main flames. Of its two frequencies,  the 223 Hz acoustic mode interacts with mainly by the convective flame dynamics along with the attached main flame and the 198 Hz mode interacts with mainly by the bulk flame dynamics close to the pilot flame. The simplified acoustic analysis conducted in this section further suggests that the 223 Hz component is likely to be the flame tube dominated mode while the 198 Hz one likely to be a plenum-flame tube coupled mode. As the flame responses to acoustic perturbations may be nonlinear and the interactions between the main and pilot stages increase the complexity, a detailed study into the nonlinear flame response would be needed in order to provide a quantitative prediction for the beating oscillations.

\section{Conclusions}

This work studies the pilot-main flame interactions in a stratified swirl burner (BASIS). Premixed methane/air mixtures with different equivalence ratios between the pilot and main stages are burned at atmospheric conditions. Three categories of tests are conducted, including the pilot flame mode, the main flame mode, and the stratified flame mode. 

In the pilot flame mode, the flame features a V-shape when the main stage is closed and an M-shape when the main stage is supplied with air. LES results show that in the case with an air stream in the main stage, the pilot stream is forced to expand more to join the main stream. The methane/air mixture from the pilot stage is therefore diluted to generate local equivalence ratio fluctuations. During combustion instabilities, a large bulk oscillation of the flame is seen. In the main flame mode, when the pilot stage is switched from closed to supplying pure air, the flame turns from an attached V-shape flame to a lifted flame. In the attached V-shape flame case, a convective motion of the flame is found during combustion instabilities. For the lifted flame, intensified unsteady heat release and a stronger combustion instability are found. 

In the stratified flame mode, beating oscillations with two close frequencies (198 Hz and 223 Hz) are found. This behaviour is found to be highly sensitive to the combination of equivalence ratios of the pilot and the main flames. Flame dynamics and acoustic-flame interactions are carefully studied by using flame oscillation sequences, Dynamic Mode Decomposition and local Rayleigh index maps. Detailed analysis to the heat release dynamics shows that the flame heat release perturbations at these two frequencies correspond to a bulk oscillation (has similarities to the unstable pilot flame case) and a convective motion (has similarities to the attached main flame case), respectively. Local Rayleigh index for each frequency confirms that acoustics at both frequencies are coupled with, i.e. driven by, their corresponding flame dynamics. A simplified acoustic model is proposed to study the frequencies of the two modes leading to the beating oscillations. This suggests that the 198 Hz mode corresponds to a plenum-flame tube coupled mode, while the 223 Hz mode corresponds to a single flame tube mode. Further study is needed to provide quantitative predictions for the beating oscillations.

In summary, this work illustrates in detail the complexity of pilot-main flame interactions and shed light on their effect on combustion instabilities in a stratified swirl burner.

\section{Acknowledgment}
This work was financially supported by the National Science and Technology Major Project (2017-III-0004-0028), National Natural Science Foundation of China (91641109, 51606004) and the European Research Council (Grant No.772080) via the ERC Consolidator Grant AFIRMATIVE (2018-23). The computational time using the CX2 HPC cluster at Imperial College is gratefully acknowledged.

\section{Supplementary materials}
\noindent (1) Videos showing flame dynamics for all of the four cases (Cases $P_fM_a$, $P_cM_f$, $P_aM_f$ and $P_fM_f$) can be found in the supplementary materials.

\noindent (2) Additional figures and tables and the associate discussion to provide further information and data.

\bibliographystyle{elsarticle-num}

\bibliography{XiaoHan.bib}

\end{document}